\documentclass[aps,prd,superscriptaddress,tightenlines]{revtex4}%{revtex4}
\input epsf
\usepackage{amsmath,amssymb,subfigure}
\usepackage{graphicx}
\usepackage{epsfig}
\usepackage{color}
\usepackage{ulem}

\newcommand{\be}{\begin{equation}}
\newcommand{\ee}{\end{equation}}

\begin{document}

%%%%%%%%%%%%%%%%%%%%%%%%%%%%%%%%%%%%%%%%%%%%%%%%%%%%%%%%%%%%%%%%%%%%%%%%%%%%%%%%%%%%%%%%%%%%  

\title{The Effects of Gravitational Slip on the Higher-order Moments of the
Matter Distribution}

\author{Scott F. Daniel\footnote{scottvalscott@gmail.com}}
\affiliation{Department of Physics and Astronomy, Dartmouth College, 
Hanover, NH 03755 USA}

\date{\today}

\begin{abstract}
%Earlier calculations in linear perturbation theory have 
%shown that departures from general relativity on cosmological
%scales (described by the model-independent parameter $\varpi$, referred to as a
%'gravitational slip') can affect the growth of structure in the universe.
Cosmological departures from general relativity 
offer a possible explanation for
the cosmic acceleration.  To linear order, these departures 
(quantified by the model-independent parameter $\varpi$, referred to as a
`gravitational slip') 
amplify or suppress
the growth of structure in the universe relative to what we would expect to see
from a general relativistic universe lately dominated by a cosmological
constant.
As structures collapse and become more dense, linear perturbation theory is an
inadequate descriptor of their behavior, and one must extend calculations to
non-linear order.  If the effects of gravitational slip extend to these higher
orders, we might expect to see a signature of $\varpi$ in the bispectrum of
galaxies distributed on the sky.  We solve the equations of motion for
non-linear perturbations in the presence of gravitational slip and find that,
while there is an effect on the bispectrum, it is too weak to be detected with
present galaxy surveys.  We also develop a formalism for incorporating scale
dependence into our description of gravitational slip.
\end{abstract}

\maketitle
\section{Introduction}
\label{intro}

A universe in which gravity obeys the laws of general relativity and is filled
with baryons and cold dark matter ought to decelerate.  
If we write the 
spatially-flat Robertson Walker metric
\begin{equation}
\label{frwmetric}
ds^2=a(\tau)^2[-d\tau^2+dr^2+r^2d\Omega^2] .
\end{equation}
deceleration means that $\dot{a}\equiv da/d\tau$
decreases with time.  This is not the case in the Universe in which we live.
Observations of distant supernovae indicate that our universe is, in fact,
accelerating ($\dot{a}$ is growing with time) 
\cite{Riess:1998cb,Perlmutter:1998np}.
To date, there are many theories vying to explain this acceleration.  
These
theories can be generally divided into two categories: theories of dark energy
and alternative theories of gravity.  If we write Einstein's equations as
$$R_{\mu\nu}-\frac{1}{2}g_{\mu\nu}R=8\pi G T_{\mu\nu},$$
dark energy attempts to explain the acceleration by
modifying the right hand side (i.e., by positing that the universe is filled
with an exotic new material).  Alternative theories of gravity attempt to
explain the acceleration by modifying the left hand side (i.e., by supposing
that the laws of gravity obey different equations of motion 
on cosmological scales than they would under general relativity).
Great diversity exists within both categories.

Dark energy can be Einstein's cosmological constant \cite{Carroll:2000fy},
a uniform energy density associated with the vacuum.  Although
this is the explanation
currently favored by the data, there is no theoretical calculation
justifying the observed value of the constant.
Dark energy may also be
a cosmological scalar field \cite{Zlatev:1998tr}
evolving slowly through its potential $V$ such that
the equation of state $w\equiv\bar{p}/\bar{\rho}<0$
at late times.
If this were the case, we would expect to see $w$ 
evolve with redshift at early times, which we do not.
Dark energy could also be a vector field 
\cite{Jimenez:2008er}, though these theories often introduce preferred reference
frames or couplings between scalar and vector perturbation modes which we have
yet to observe.

Alternative gravity can be a scalar-tensor theory \cite{Schimd:2004nq}
in which the dark energy scalar field is non-minimally coupled to the
curvature terms in the Einstein-Hilbert action.  These theories tend to predict
departures from Newton's law of gravitation, though the simplest of them
merely manifest themselves as a rescaling of Newton's constant $G$.
There are also theories which modify gravity by introducing an arbitrary
function of the Ricci scalar $f(R)$ into the gravitational Lagrangian
\cite{Carroll:2004de,Acquaviva:2004fv,Zhang:2005vt}.  These theories tend to
introduce new scale-dependent effects into the evolution of large scale
structure.
Tensor-vector-scalar (TeVeS) theory
\cite{Bekenstein:2004ne,Skordis:2005eu}
adds tensor and vector fields to the mix of scalar-tensor theory.  These theories
require precise couplings to avoid the pit-falls of vector dark energy noted
above.
Multi-dimensional 
braneworld theories (like the
Dvali-Gabadadze-Porrati -- DGP -- model)
\cite{Dvali:2000hr,Lue:2005ya,Song:2006jk} attempt to account for cosmic
acceleration by allowing gravity to act in dimensions outside of our $3+1$
Universe, imposing a scale beyond which the expected gravitational attraction is
damped.
Theories inspired by quantum mechanics, such as a Lorentz-violating 
massive graviton
\cite{Bebronne:2007qh}, reproduce solar system tests of gravity but introduce new
classes of cosmological perturbations which may or may not show up in our
measurements of large scale structure, depending on initial conditions.  

All of these theories have solutions that can
provide late-time cosmic acceleration.  Successfully navigating this
labyrinth of viable theories requires a set of observations complementary 
to the
expansion of the universe for 
which different theories of gravity or dark energy
predict different effects.
The growth of cosmic structure provides one such set of observations.
We model that growth using cosmic perturbation theory.
If we consider only perturbations that transform like scalars, the first-order
perturbed form of equation (\ref{frwmetric}) is
\begin{equation}
\label{perturbedmetric}
ds^2=a^2[-(1+2\psi)d\tau^2+(1-2\phi)d\vec{x}^2] .
\end{equation}
Neglecting the cosmic scale factor $a$, 
the potentials $\psi$ and $\phi$ -- known 
respectively as the Newtonian and
longitudinal potentials --  supply the right hand sides of Newton's law
of gravitation $\ddot{\vec x}=-\vec\nabla\psi$ and the Poisson equation 
$4\pi Ga^2\delta\rho=\nabla^2\phi$.  Consistent with Newtonian dynamics, 
general relativity in the presence of non-relativistic
stress-energy predicts that $\phi=\psi$.
Alternative theories of gravity make no such
guarantee.  Each of the gravity theories cited 
above implies its own unique relationship
between $\phi$ and $\psi$
\cite{Schimd:2004nq,Acquaviva:2004fv,Zhang:2005vt,Skordis:2005eu,Dvali:2000hr,Lue:2005ya,Song:2006jk,Bebronne:2007qh}.

Lacking theoretical justification to prefer one 
alternative gravity theory over another,
we will content ourselves searching for evidence of gravitational slip 
(our word for the difference between $\phi$ and $\psi$) in
general.
Following Caldwell {\it et al}. \cite{Caldwell:2007cw}, we parametrize the
relationship between $\phi$ and $\psi$ as
\begin{eqnarray}
\psi&=&(1+\varpi)\phi\label{parametrization}\\
\varpi&=&\varpi_0a^{3} .\label{varpieqn}  
\end{eqnarray}
Obviously, the scale independence and redshift dependence of equation
(\ref{varpieqn}) are assumptions on our part.
Because we are interested in gravity theories which might explain the late time
cosmic acceleration, we suppose that $\phi\approx\psi$ at 
high redshift and that gravitational slip should grow as the inverse of the
matter density.
For the purposes of our calculations, we will
assume that the expansion history of the Universe
exactly matches
that of a $\Lambda$CDM universe, that is
a universe which obeys general relativity ($\varpi$=0)
and in which the acceleration is caused by a cosmological 
constant contributing a fraction
$\Omega_\Lambda<1$ to the current density of the Universe.  

Equation (\ref{parametrization}) 
is obviously not the only possible parametrization of modified gravity.
References \cite{Bertschinger:2006aw,Zhang:2007nk,Hu:2007pj,Amendola:2007rr,Jain:2007yk,Zhang:2008ba,Bertschinger:2008zb,Hu:2008zd}
all explore different model-independent expressions of $\phi\ne\psi$.  
Reference \cite{Daniel:2008et} discusses the consistency of our choice
(\ref{parametrization}) with these other parametrizations.  In the end, it is
simply a question of nomenclature.

References \cite{Daniel:2008et,Daniel:2009kr}, 
derived the linear-order equations of
motion by perturbing a Robertson Walker metric (\ref{frwmetric}) in
the presence of a homogeneous, isotropic stress energy tensor
\begin{equation}
\label{cosmoTmunu}
T^\mu_\nu=\text{diag}(-\bar\rho,\bar{p},\bar{p},\bar{p})
\end{equation}
and constrained $\varpi_0$ against CMB, supernova, and weak lensing (of
galaxies) 
data sets. 
Specifically, reference \cite{Daniel:2009kr} found the constraint
$\varpi_0 =0.09{}^{+0.74}_{-0.59}\, (2\sigma)$.
Reference \cite{Serra:2009kp} also tested equation (\ref{parametrization}),
but against CMB and weak lensing (of the CMB) data sets.  They also 
promoted the
redshift dependence of equation (\ref{varpieqn}) to a new free parameter.
They found $\varpi_0=1.67^{+3.07}_{-1.87}\, (2\sigma)$.  Both results are
consistent with a $\Lambda$CDM universe obeying the laws of general relativity,
but with significant room for departure ($\phi=\psi$ within
a factor of a few).  In this paper, we will attempt to complement those
explorations by calculating the effect of non-zero $\varpi$ on the
growth of structure in the Universe beyond linear order.
%\cite{Caldwell:2007cw} \cite{Daniel:2008et,Daniel:2009kr}

As gravitational collapse of structures progresses, the gravitational fields
within overdense regions of the universe grow so that effects beyond linear
order become important.  Even in a $\varpi=0$ universe, this causes the
distribution of overdensities to evolve away from their initial Gaussian
distribution \cite{Peebles:1980}.  These departures from Gaussianity are
evidenced in the bispectrum, the Fourier transform of the three-point
correlation function of galaxies.
By altering the growth of structure
\cite{Daniel:2009kr}, $\varpi$ ought also to alter these departures from
gaussianity, an effect which we hope will be detectable
in modern galaxy surveys.  Section \ref{eomsectionskew}
will derive the equations of motion for cosmic overdensities in the case of
$\varpi\ne0$ to ``quasi-linear'' order.  Section \ref{fouriersection} will
translate these results into Fourier space.  Section \ref{bispsection} will
calculate the resulting effects on the bispectrum.  Section \ref{redshiftsection} includes the effects of
galaxy bias and redshift distortions.  Section
\ref{scalesection} will consider the effect of adding scale dependence to
$\varpi$.  Section \ref{conclusion} will discuss our results.  Appendix
\ref{lptsection} compares the results of our equations of motion (derived in
Section \ref{eomsectionskew} using Eulerian Perturbation Theory) to previous
results derived in Lagrangian Perturbation Theory.  
Appendix \ref{skewnesssection} discusses the averaged second order moment of the overdensity distribution (the skewness).
Appendix
\ref{kurtosissection} extends our results to the next higher quasi-linear order,
calculating the effect of scale-independent $\varpi$ on the kurtosis of the
overdensity distribution.

%%%%%%%%%%%%%%%%%%%%%%%%%%%%%%%%%%%%%%%%%%%%%%%%%
\section{Equations of motion}
\label{eomsectionskew}

This paper will work in the convention of Eulerian Perturbation Theory (see
Appendix \ref{lptsection} for a discussion of Lagrangian Perturbation Theory).
We will directly evolve the perturbed quantities $\phi$,
$\delta\equiv(\rho-\bar\rho)/\bar\rho$ and $\vec{v}$, the perturbed velocity
field in the cold dark matter fluid.
References \cite{Caldwell:2007cw,Daniel:2008et,Daniel:2009kr} discuss how to 
modify the linear-order equations of motion for these perturbed quantities in 
the case of $\varpi\ne0$.  For the sake of brevity, we will merely summarize 
that discussion here.  Of the linear-order perturbed Einstein equations 
presented in reference \cite{Ma:1995ey}, the time-time and diagonal 
space-space are discarded.  The time-space Einstein equation is preserved 
as a consequence of the assumption
that the cold dark matter fluid remains, on average, at rest in our reference
frame.  A term is added to the off-diagonal space-space Einstein equation so 
to provide for $\varpi$ as a new source of cosmic shear.  Because we have
discarded the time-time and diagonal space-space Einstein equations, we can no
longer assume that the Poisson equation is valid.  This will be important when
deriving the quasi-linear equations of motion below.

At quasi-linear order, gravitational collapse has advanced such that
\begin{equation}
\label{quasilinearassumption}
\delta,v/c\gg \phi/c^2,\psi/c^2, 
\end{equation}
i.e.
the dynamics of the local fluid dominate the dynamics of the
spacetime.  
The metric is still Robertson Walker, but now, the stress
energy tensor (\ref{cosmoTmunu}) is replaced by that of a perfect fluid
$$T^{\mu\nu}=(\bar\rho+\bar{p})u^\mu u^\nu+g^{\mu\nu}\bar{p}$$
where $u^\mu$ are the components of the fluid's four-velocity
\cite{Schutz:1985jx}.  To consistently determine the smallness of
perturbed quantities, we explicitly include factors of $c$ in the
perturbed metric (\ref{perturbedmetric}), giving
\begin{equation}
\label{perturbedmetricskew}
\frac{ds^2}{c^2}=-a^2\left(1+2\frac{\psi}{c^2}\right)d\tau^2+
\frac{a^2}{c^2}\left(1-2\frac{\phi}{c^2}\right)d\vec{x}^2
\end{equation}
\noindent
We derive our equations of motion by solving for the dynamics of this
reformulated system to zeroth order in $1/c$.  This is equivalent to
assumption (\ref{quasilinearassumption}).

Requiring $u^\mu u_\mu=-1$, we find that
\begin{equation}
\label{fourvelocity}
u^\mu=\gamma\left[\frac{1}{a}\left(1-\frac{\psi}{c^2}\right),
\frac{\vec{v}}{a}\left(1+\frac{\phi}{c^2}\right)\right]
\end{equation}
\noindent
where $\gamma=1/\sqrt{1-\vec{v}^2/c^2}$, as usual.  
Thus, to the required order (because we are considering
perturbations in the matter distribution, we set $\bar{p}=0$),
\begin{eqnarray}
\nabla_\mu T^{\mu
0}&=&
\frac{\gamma\bar\rho}{a^2}
\left(\dot{\delta}+\partial_i(1+\delta)v^i\right)=0\label{deltadoteqn}\\
\nabla_\mu T^{\mu i}&=&\frac{\gamma\bar{\rho}}{a^2}
\left(\dot{v}^i+\mathcal{H}v^i+v^j\partial_jv^i+\partial_i\psi\right)
=0\label{vdoteqn}\\
R_{0i}&-&\frac{1}{2}g_{0i}R=8\pi GT_{0 i}\nonumber\\
&&\phantom{\frac{1}{2}g_{0i}R}=\frac{1}{c^2}
\left(2\partial_{0i}\phi+2\mathcal{H}\partial_i\psi\right)\nonumber\\
&&\phantom{\frac{1}{2}g_{0i}R}=-8\pi G
a^2v^i\bar{\rho}(1+\delta)\frac{1}{c^2}\label{einstein0i}
\end{eqnarray}
\noindent
where $\mathcal{H}$ is the conformal time Hubble parameter
$\mathcal{H}\equiv\dot{a}/a$.
Note that, to lowest order, $\partial^i=a^{-2}\delta^{ij}\partial_j$.
Equations (\ref{deltadoteqn})
and (\ref{vdoteqn}) also correspond to 
equations (2) and (3) of Catelan and Moscardini's paper deriving the
quasi-linear fluid equations of motion in unmodified general relativity
\cite{Catelan:1993zf}.  Their third equation of motion derives from the
Poisson equation, which we discard in favor of our equation
(\ref{einstein0i}).
(Note that Catelan and Moscardini use the coordinate time $t$ where
we use the conformal time $\tau$).  
Following Catelan and Moscardini's lead, we substitute
equation (\ref{deltadoteqn}) into the divergence of 
equation (\ref{vdoteqn}) to
get
\begin{eqnarray}
\ddot{\delta}+\mathcal{H}\dot\delta
&=&(1+\varpi)\partial_i\delta\partial_i\phi\label{eom}\\
&&+
\partial_{ij}[(1+\delta)v^iv^j]+(1+\delta)
(1+\varpi)\nabla^2\phi\nonumber
\end{eqnarray}
\noindent
where we have used equation (\ref{parametrization}) to rewrite $\psi$ in terms
of $\phi$ and $\varpi$.
In our notation $\nabla^2\equiv\partial_i\partial_i$.
Using equation (\ref{deltadoteqn}), we can rewrite equation
(\ref{einstein0i}) as
\begin{equation}
\label{einsteinrewrite1}
\frac{3}{2}\mathcal{H}^2\Omega_m\dot{\delta}=
\nabla^2\left(\dot{\phi}+\mathcal{H}(1+\varpi)\phi\right)
\end{equation}
\noindent
the time derivative of which is
\begin{eqnarray}
\frac{3}{2}\Omega_m\mathcal{H}^2\left(\ddot{\delta}
-\mathcal{H}\dot{\delta}\right)&=&\nabla^2
\bigg(\ddot{\phi}+\mathcal{H}(1+\varpi)\dot{\phi}\nonumber\\
&&+\mathcal{H}^2(1+\varpi)\phi(1-\frac{3}{2}\Omega_m)+
\mathcal{H}\dot{\varpi}\phi\bigg) .\nonumber
\end{eqnarray}
Using equation (\ref{einsteinrewrite1}), we find
\begin{eqnarray}
\frac{3}{2}\Omega_m\mathcal{H}^2\left(\ddot{\delta}+
\mathcal{H}\dot{\delta}\right)&=&\nabla^2\bigg(\ddot{\phi}+
\mathcal{H}(3+\varpi)\dot{\phi}\label{einsteinrewrite}\\
&&+\mathcal{H}^2(1+\varpi)\phi(3-\frac{3}{2}\Omega_m)+
\mathcal{H}\dot{\varpi}\phi\bigg) .\nonumber
\end{eqnarray}
Equations (\ref{eom}) and (\ref{einsteinrewrite}) 
provide an algorithm by which we can solve
for $\phi$ and $\delta$ to arbitrary order.  
Once we have solved for $\phi$ and $\delta$,
it is a simple matter to use equation (\ref{deltadoteqn}) 
to find $\vec{v}$.

Assume that $\phi$, $\delta$ and $\vec{v}$ can be 
expanded as $\phi=\sum_i\phi^{(i)}$ etc., where
$\phi^{(i)}\gg\phi^{(j>i)}$ (i.e., ``$\phi^{(i)}$ is 
the $i$th order part of $\phi$'').  In that case,
we can expand equations (\ref{eom}) and (\ref{einsteinrewrite}) to 
a given order $n$, then use
the proportionality between their left hand sides 
to get a single equation of the form
\begin{eqnarray}
\nabla^2\Bigg(\ddot{\phi}^{(n)}&+&\dot\phi^{(n)}\mathcal{H}(3+\varpi)
%&&
+\phi^{(n)}\left[\mathcal{H}^23(1+\varpi)(1-\Omega_m)
+\mathcal{H}\dot{\varpi}\right]\Bigg)%\nonumber\\
%&&\qquad\qquad
=\frac{3}{2}\Omega_m\mathcal{H}^2S^{(n)}\label{delsquaredphi}\\
S^{(n)}&\equiv&\sum_{a+b+c=n}\Big\{
(1+\varpi)\partial_i\delta^{(a)}\partial_i\phi^{(b)}%\nonumber\\
%&&\phantom{\sum}
+\partial_{ij}[(1+\delta^{(a)})v^{(b)i}v^{(c)j}]%\nonumber\\
%&&\phantom{\sum}
+\delta^{(a)}(1+\varpi)\nabla^2\phi^{(b)}\Big\}
\label{sourceterm}
\end{eqnarray}
where the source terms $S^{(n)}$ come from the non-linear 
part of equation (\ref{eom}).
After solving equation (\ref{delsquaredphi}) for 
$\nabla^2\phi^{(n)}$, one can use
equation (\ref{eom})
\begin{equation}
\label{deltaddoteqn}
\ddot{\delta}^{(n)}+\mathcal{H}\dot{\delta}^{(n)}=S^{(n)}+
(1+\varpi)\nabla^2\phi^{(n)}
\end{equation}
\noindent
to solve for $\delta^{(n)}$.

Comparing equations (\ref{delsquaredphi}) and (\ref{deltaddoteqn}),
we see that, even though the Poisson 
equation no longer holds, $\phi$ and $\delta$ are 
still separable at first order ($\phi=f(\tau)\varphi(\vec{x})$ and 
$\delta=D(\tau)\xi(\vec{x})$) and related such that
the spatial parts of $\delta$ are the Laplacians 
of the spatial parts of $\phi$ (i.e., $\xi=\nabla^2\varphi$).  
Indeed, to first order
\begin{eqnarray}
\phi^{(1)}&=&f(\tau)\varphi(\vec{x})\label{phifirstorder}\\
\delta^{(1)}&=&D(\tau)\nabla^2\varphi(\vec{x})\label{deltafirstorder}\\
\vec v^{(1)}&=&-\dot D\vec\nabla\varphi\label{vfirstorder}\\
\dot{f}+f\mathcal{H}(1+\varpi)&=&\frac{3}{2}\Omega_m\mathcal{H}^2\dot{D}\nonumber\\
\ddot{D}+\mathcal{H}\dot{D}&=&(1+\varpi)f .\nonumber
\end{eqnarray}
At higher orders, the expressions are less compact
\begin{eqnarray}
\phi^{(2)}&=&\alpha(\tau)A(\vec x)+\beta(\tau)B(\vec x)\nonumber\\
\delta^{(2)}&=&\mathcal{D}^{a}\nabla^2A+\mathcal{D}^{b}\nabla^2B\label{deltasecondorder}\\
\vec v^{(2)}&=&\vec\nabla A(-\dot{\mathcal{D}}^{a}+D\dot{D})
-\vec\nabla B\dot{\mathcal{D}}^b ,
\label{vsecondorder}
\end{eqnarray}
where
\begin{eqnarray}
\nabla^2A&=&\partial_i\left(\nabla^2\varphi\partial_i\varphi\right)\label{Aeqn}\\
\nabla^2B&=&\partial_{ij}\left(\partial_i\varphi\partial_j\varphi\right)\label{Beqn}\\
\ddot{\alpha}&=&-\dot{\alpha}\mathcal{H}(3+\varpi)
+\frac{3}{2}\Omega_m\mathcal{H}^2(1+\varpi)Df
-\alpha\left[3\mathcal{H}^2(1+\varpi)(1-\Omega_m)+\dot{\varpi}\mathcal{H}\right]\label{alphaeqn}\\
\ddot{\beta}&=&-\dot{\beta}\mathcal{H}(3+\varpi)
+\frac{3}{2}\Omega_m\mathcal{H}^2\dot{D}^2
-\beta\left[3\mathcal{H}^2(1+\varpi)(1-\Omega_m)+\dot{\varpi}\mathcal{H}\right]\label{betaeqn}\\
\ddot{\mathcal{D}}^{a}&=&-\mathcal{H\dot{D}}^{a}+(1+\varpi)\alpha+(1+\varpi)Df\label{curlyD1}\\
\ddot{\mathcal{D}}^{b}&=&-\mathcal{H\dot{D}}^{b}+
(1+\varpi)\beta+\dot{D}^2 .\label{curlyD2}
\end{eqnarray}
In the $\varpi=0$ limit, these expressions agree 
with the GR results of \cite{Kamionkowski:1998fv}.
There are seven different spatial functions comprising $\phi^{(3)}$.  
We present them in Appendix \ref{eomappendix}.

\begin{figure}[!h]
\subfigure[]{
\includegraphics[scale=0.3]{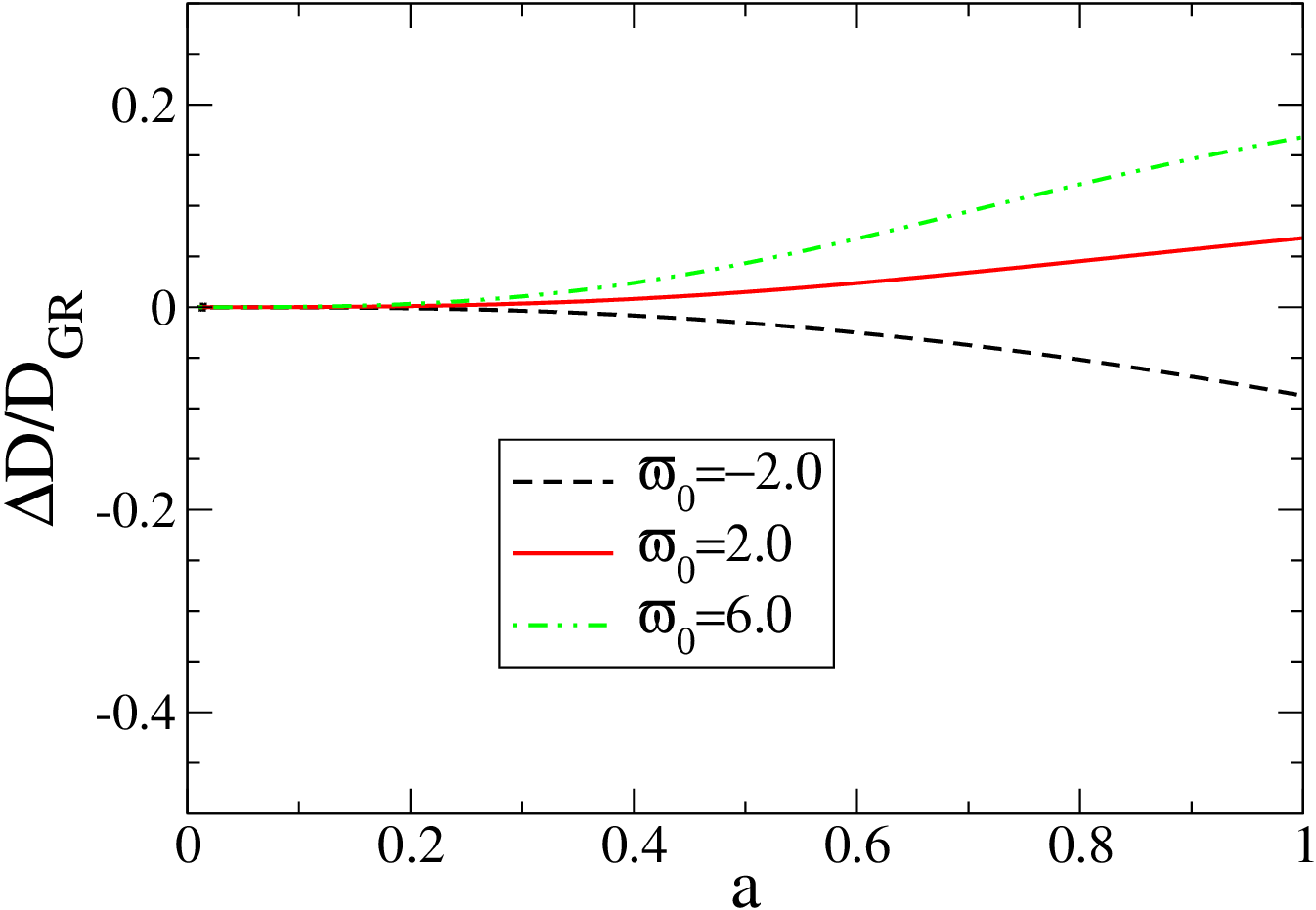}
\label{D_1growthfig}
}
\subfigure[]{
\includegraphics[scale=0.3]{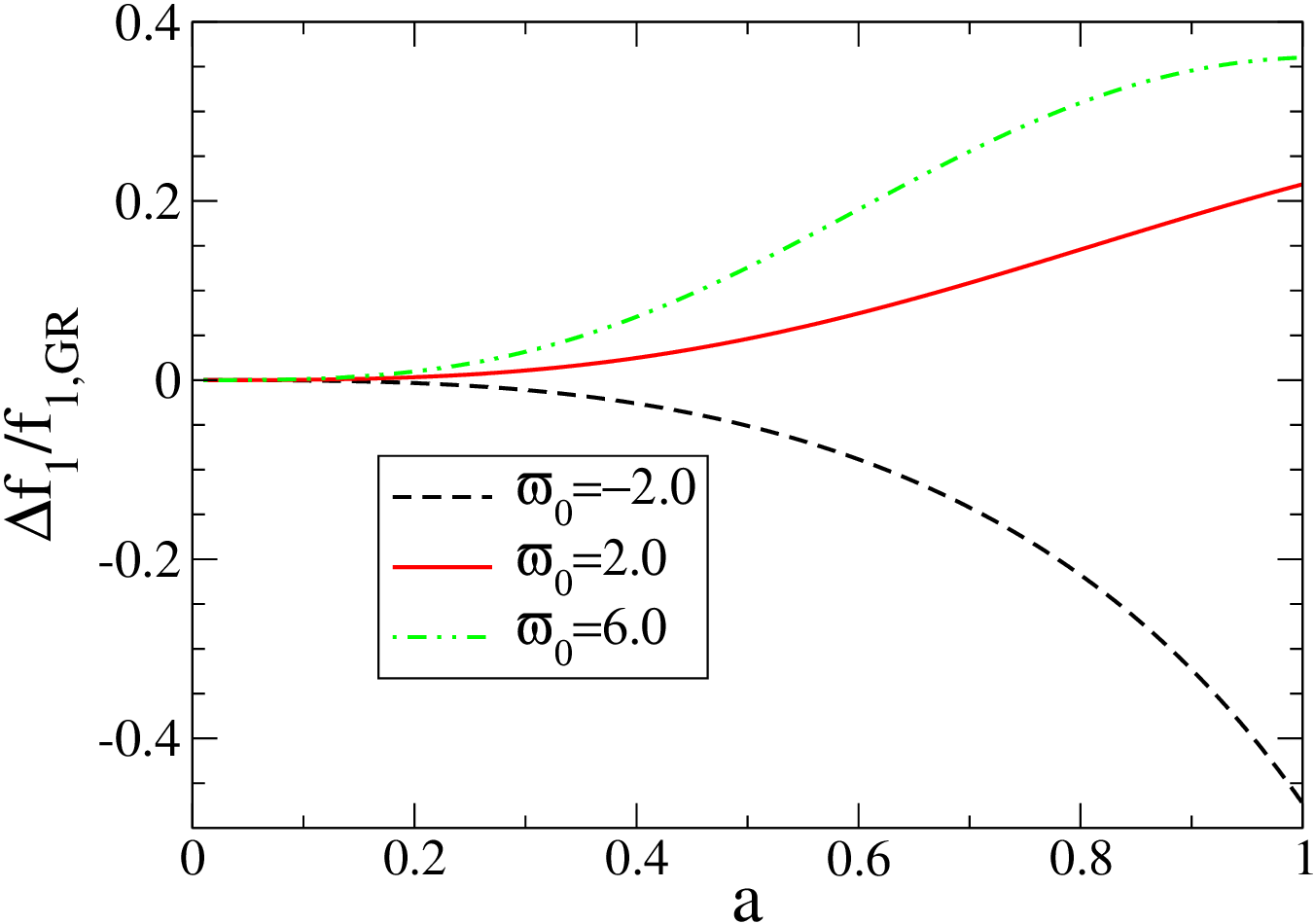}
\label{f1growthfig}
}
\caption{We plot the change in the first order growth functions
$D$ and $f_1\equiv \frac{a}{D}\frac{dD}{da}$
resulting from varying $\varpi_0$.  Note that $f_1$ is a different parameter
from the $f$ used in equation (\ref{phifirstorder}).  
All other parameters are 
set to be the WMAP 5-year maximum likelihood values
\cite{Dunkley:2008ie}.
The change is calculated relative to a $\varpi=0$ (GR) model so that
$\Delta D/D_{GR}=(D-D_{GR})/D_{GR}$.  As found in \cite{Daniel:2008et},
$\varpi_0>0$ amplifies the growth of $\delta^{(1)}$, while $\varpi_0<0$
suppresses it.
}
\label{growth1fig}%
\end{figure}

\begin{figure}[!h]
\subfigure[]{
\includegraphics[scale=0.3]{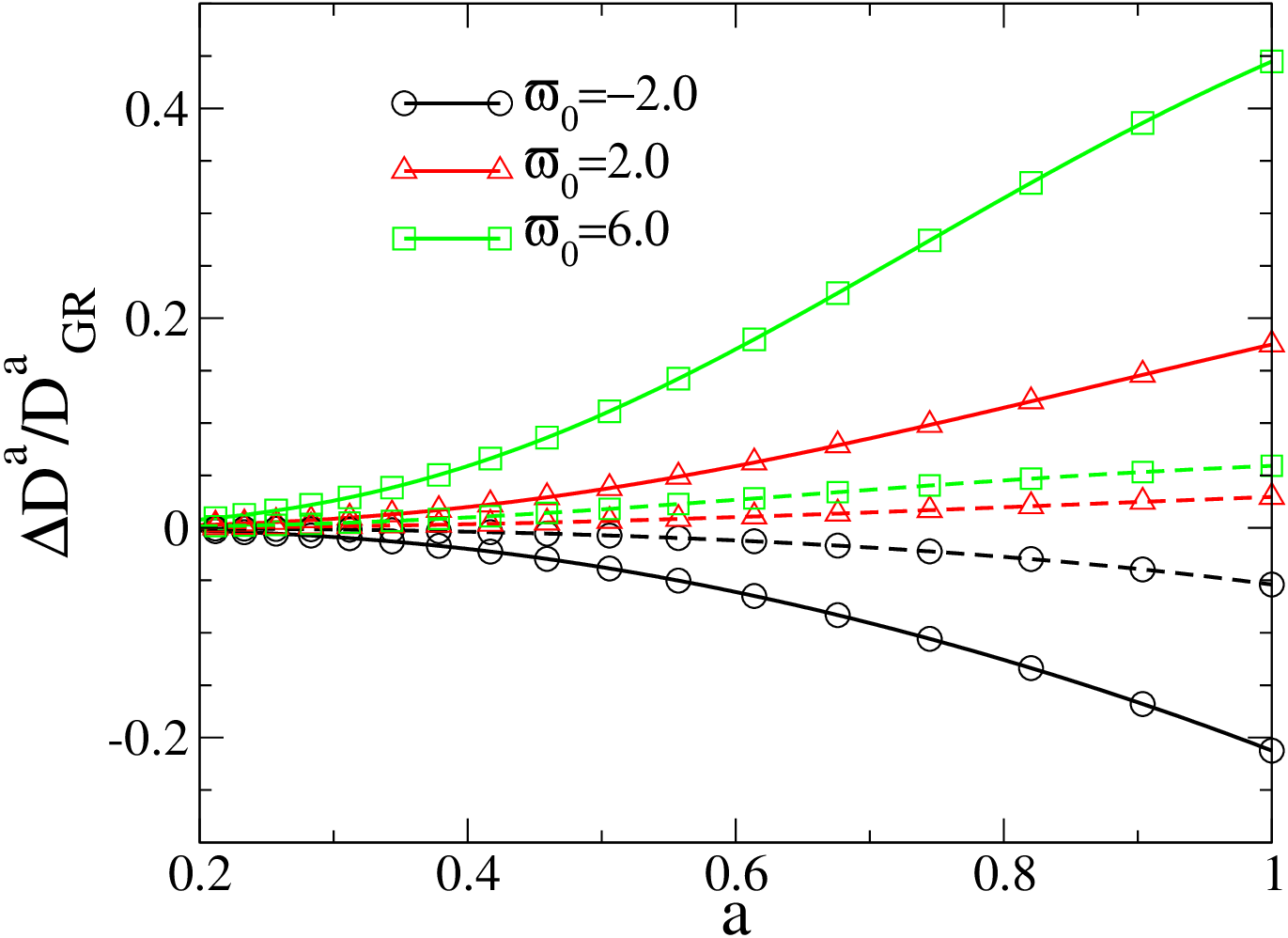}
\label{D1growthfig}
}
\subfigure[]{
\includegraphics[scale=0.3]{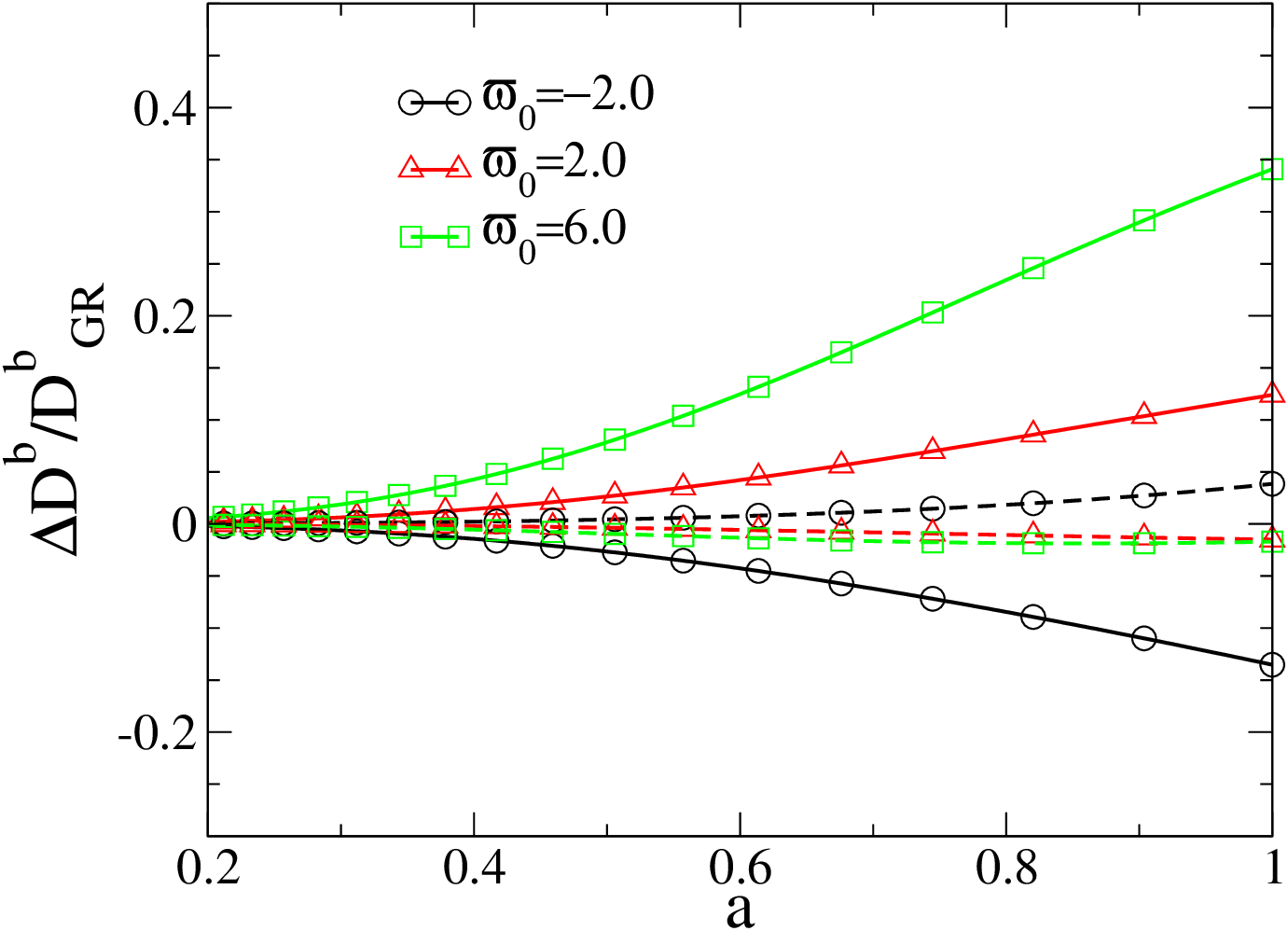}
\label{D2growthfig}
}
\caption{We plot the change in the second order growth functions
$\mathcal{D}^a$ and $\mathcal{D}^b$
resulting from varying $\varpi_0$.  
All other parameters are 
set to be the WMAP 5-year maximum likelihood values
\cite{Dunkley:2008ie}.
In the case of solid lines,
$\Delta\mathcal{D}^i\equiv\mathcal{D}^i-\mathcal{D}^i_{GR}$.
In the case of dashed lines, the second order growth functions are normalized
by a factor of $D^2$ ($D$ is the first order growth function) so that
$\Delta\mathcal{D}^i\equiv\mathcal{D}^i(D_{GR}/D)^2-\mathcal{D}^i_{GR}$.
The normalization greatly diminishes the effect of varying $\varpi_0$,
indicating that most of the effect of $\varpi_0$ on the higher order growth
functions enters as a normalization.
}
\label{growth2fig}%
\end{figure}

We now only lack initial conditions on the relevant growth terms 
in our attempt to solve
for $\delta$ in the quasi-linear regime.
Since the power spectrum $P_\delta$ is initially Gaussian, 
non-linear growth terms (i.e., $\alpha$ and $\beta$ from 
equation (\ref{deltasecondorder})) are integrated from the 
early-time initial conditions 
$\alpha_i=0$, $\dot{\alpha}_i=0$.  To find the initial 
conditions for the linear growth terms $f$ and $D$, we 
combine the first order equations of motion (\ref{eom}) and 
(\ref{einsteinrewrite1}), giving
\begin{eqnarray}
\ddot{D}+\dot{D}\mathcal{H}&=&(1+\varpi)f\nonumber\\
\dot{f}+f\mathcal{H}(1+\varpi)&=&\frac{3}{2}\Omega_m\mathcal{H}^2\dot{D}\nonumber
\end{eqnarray}
to get
\begin{eqnarray}
\dot{f}&=&\frac{3}{2}\Omega_m\mathcal{H}^2\dot{D}-\mathcal{H}^2\dot{D}-\mathcal{H}\ddot{D}\nonumber\\
&=&-\dot{\mathcal{H}}\dot{D}-\mathcal{H}\ddot{D}\nonumber\\
&=&-\frac{d}{d\tau}\left(\mathcal{H}\dot{D}\right)\nonumber\\
f&=&-\mathcal{H}\dot{D}+C\label{fconst}
\end{eqnarray}
where $C$ is a constant.  Since we assume that 
$\lim_{a\rightarrow0}\varpi=0$,
we use the $\Lambda$CDM equations of motion combined with the Poisson-equation result
$f=(3/2)\Omega_m\mathcal{H}^2D$ to 
find the correct value for $C$ for a given $\Omega_m$.  
This value changes depending 
on the initial conditions assumed for $D$ and $\dot{D}$.  
However, the normalization of the measured statistic
(\ref{qdef}) means that our final results are resilient 
to such choices.  Figure \ref{growth1fig} plots the effect of gravitational slip
on the first order growth function $D$ and its derivative.  Figure
\ref{growth2fig} plots the same effects on the second order growth functions
$\mathcal{D}^a$ and $\mathcal{D}^b$.

%third order phi eom is generalized from pattern (see "once and for all" derivation of Ddotdot eqn in terms of Dprimeprime in notebook 3)

%third order delta comess from eom; note that the G source terms are the same as those in third order hi eqns; just have to add phi terms from the del square phi to the three that enters in

%%%%%%%%%%%%%%%%%%%%%%%%%%%%%%%%%%%%%%

\section{Fourier transform of $\langle\delta^{n}\rangle$}
\label{fouriersection}

In Section \ref{bispsection} we will derive an expression for the bispectrum 
of the matter distribution in the case of
$\varpi\ne0$.  This calculation will require taking the Fourier transform of
terms like $\langle\delta^{n}(\vec{x})\rangle$ for $n>1$, 
which we illustrate below. 

If we expand $\delta(\vec{x})=\sum_{i}\delta^{(i)}(\vec{x})$, then
\begin{eqnarray}
\langle\delta^n\rangle&=&\langle(\delta^{(1)})^n\rangle
+n\langle\delta^{(2)}(\delta^{(1)})^{n-1}\rangle
+n\langle\delta^{(3)}(\delta^{(1)})^{n-1}\rangle%\nonumber\\
%&&
+\binom{n}{2}\langle(\delta^{(2)})^2(\delta^{(1)})^{n-2}\rangle+\dots
\label{deltatothen}
\end{eqnarray}
(note: superscripts in parentheses are orders of expansion; superscripts
outside parentheses are exponents).  
Therefore, in order to get
the Fourier tranform of $\langle\delta^n\rangle$, 
we first need the Fourier
transform of $\langle\delta^{(i)}\rangle$.  From the form of equation
(\ref{deltaddoteqn}), we can see that  
\begin{equation}
\label{deltageneralorder}
\delta^{(i)}(\vec{x})=D^{(i)}(\tau)
\Delta(\varphi_1,\varphi_2,\dots,\varphi_i)
\end{equation}
where $D^{(i)}$ stands for some combination of $i$th order growth
functions and $\Delta$ is some combination of spatial derivatives
acting on the $i$ powers of $\varphi(\vec{x})$
(see, for example, equation \ref{deltasecondorder}).
Taking the Fourier transform of this qualitative $\delta^{(i)}$,
we find (here we will introduce the notation
$\tilde\delta^{(i)}_a\equiv\tilde\delta^{(i)}(\vec{k}_a)$)
\begin{eqnarray}
&&\tilde\delta^{(i)}(\vec{k})=\int\frac{d^3x}{(2\pi)^{3/2}}
e^{-i\vec{k}\vec{x}}
D^{(i)}\Delta(\varphi_1,\dots,\varphi_i)\label{firstfourier}\\
&&=\int\frac{d^3xd^3k_{1\dots i}}{(2\pi)^{\frac{3}{2}(i+1)}}
e^{i\vec{x}(\sum_{j=1}^i\vec{k}_j-\vec{k})}
D^{(i)}\mathcal{K}
\prod_{m=1}^i\tilde\varphi_m\label{secondfourier}\\
&&=\int\frac{d^3k_{1\dots i}}{(2\pi)^{\frac{3}{2}(i-1)}}
\delta^3_D(\vec{k}-\sum_{j=1}^i\vec{k}_j)
D^{(i)}\mathcal{K}
\prod_{m=1}^i\tilde\varphi_m\label{thirdfourier}\\
&&=\int\frac{d^3k_{1\dots i}}{(2\pi)^{\frac{3}{2}(i-1)}}
\delta^3_D(\vec{k}-\sum_{j=1}^i\vec{k}_j)
%\nonumber\\
%&&\qquad\qquad
\times\frac{(-1)^iD^{(i)}\mathcal{K}}{(D^{(1)})^i}
\prod_{{l}=1}^i\frac{\tilde\delta^{(1)}_{l}}{k^2_l}
\label{fourthfourier}
\end{eqnarray}
(note $d^3k_{1\dots i}\equiv\prod_{j=1}^id^3k_j$)
where, in going from equation (\ref{firstfourier}) to
(\ref{secondfourier}), we have taken the Fourier transform of each
individual factor of $\varphi(\vec{x})$ in equation
(\ref{deltageneralorder}).  $\mathcal{K}$ represents the
combination of wave vectors
$\{\vec{k}_1,\vec{k}_2,\dots\vec{k}_i\}$ deriving from 
the differential operator $\Delta$.  $\delta^3_D$ is a Dirac delta function.
We used equation
(\ref{deltafirstorder}) to go from equation (\ref{thirdfourier}) to
(\ref{fourthfourier}).

From equation (\ref{fourthfourier}), we can see that a term of the form
$\delta^{(i)}(\vec{x})(\delta^{(1)}(\vec{x}))^n$ can be
written in terms of Fourier transforms as
\begin{eqnarray}
\delta^{(i)}(\vec{x})(\delta^{(1)}(\vec{x}))^n&=&
\int\frac{d^3kd^3k_{1\dots n}}{(2\pi)^{\frac{3}{2}(n+1)}}
e^{i\vec{x}(\vec{k}+\sum_{j=1}^n\vec{k}_j)}
%\nonumber\\
%&&\phantom{\int\frac{d^3kd^3k_{1\dots n}}{(2\pi)^{\frac{3}{2}(n+1)}}
%e^{i\vec{x}(\vec{k}+\sum_{j=1}^n\vec{k}_j)}}
\times\tilde\delta^{(i)}\prod_{m=1}^n\tilde\delta^{(1)}_m\nonumber\\
&=&\int\frac{d^3kd^3k_{1\dots n}}{(2\pi)^{\frac{3}{2}(n+1)}}
\frac{d^3k^\prime_{1\dots i}}{(2\pi)^{\frac{3}{2}(i-1)}}
 e^{i\vec{x}(\vec{k}+\sum_{j=1}^n\vec{k}_{j})}
\prod_{m=1}^n\tilde\delta^{(1)}_m%\nonumber\\
%&&\qquad
\times \delta^3_D(\vec{k}-\sum_{l=1}^i\vec{k}^\prime_l)
\frac{(-1)^iD^{(i)}}{(D^{(1)})^i}
\mathcal{K}
\prod_{b=1}^i\frac{\tilde\delta^{(1)\prime}_b}{k^{\prime 2}_b}
\label{compoundfourier}
\end{eqnarray}
where, to be explicit, the $k^\prime_j$ wave vectors come from the
Fourier transform $\tilde{\delta}^{(i)}$.  
Integrating the right hand
side of equation (\ref{compoundfourier}) over the wave vector
$\vec{k}$, we see that
\begin{eqnarray}
\delta^{(i)}(\vec{x})(\delta^{(1)})^n&=&
\int\frac{d^3k_{1\dots i+n}}{(2\pi)^{\frac{3}{2}(i+n)}}
e^{i\vec{x}\cdot(\sum_{j=1}^{i+n}\vec{k}_j)}%\nonumber\\
%&&\qquad
\times\frac{D^{(i)}}{(D^{(1)})^i}
\frac{\mathcal{K}}{\prod_{l=1}^ik^2_l}
\prod_{j=1}^{i+n}\tilde\delta^{(1)}_j
\nonumber
\end{eqnarray}
whence
\begin{eqnarray}
\langle\delta^{(i)}(\vec{x})(\delta^{(1)})^n\rangle&=&
\int\frac{d^3k_{1\dots i+n}}{(2\pi)^{\frac{3}{2}(i+n)}}
e^{i\vec{x}\cdot(\sum_{j=1}^{i+n}\vec{k}_j)}%\nonumber\\
%&&\phantom{\int d^3k}
\times\frac{(-1)^iD^{(i)}}{(D^{(1)})^i}
\frac{\mathcal{K}}{\prod_{l=1}^ik^2_l}%\nonumber\\
%&&\phantom{\int d^3k}
\times\langle\tilde\delta^{(1)}_1
\dots\tilde\delta^{(1)}_{i+n}\rangle .
\label{fourieravg}
\end{eqnarray}
At this point, it is useful to recall that, for a Gaussian field (like
$\tilde\delta^{(1)})$,
\begin{eqnarray}
\langle\tilde\delta^{(1)}_1\tilde\delta^{(1)}_2\dots\tilde\delta^{(1)}_n\rangle&=&
0\phantom{\langle\delta^1\delta\delta\dots \delta\dots2
+\text{permutations}}\qquad\text{$n$
odd}\label{gaussianodd}\\
&=&\langle\tilde\delta^{(1)}_1\tilde\delta^{(1)}_2\rangle\langle\tilde\delta^{(1)}_3\tilde\delta^{(1)}_4\rangle\dots\langle\tilde\delta^{(1)}_{n-1}\tilde\delta^{(1)}_n\rangle\nonumber\\
&&\phantom{\langle\delta^1\delta\delta\dots \delta\dots2}
+\phantom{2}\text{permutations}\qquad\text{$n$ even}\label{gaussianeven}
\end{eqnarray}
so that, for example
\begin{eqnarray}
\langle\tilde\delta^{(1)}_1\tilde\delta^{(1)}_2\tilde\delta^{(1)}_3\rangle&=&0\nonumber\\
\langle\tilde\delta^{(1)}_1\tilde\delta^{(1)}_2\tilde\delta^{(1)}_3\tilde\delta^{(1)}_4\rangle&=&
\phantom{+}\langle\tilde\delta^{(1)}_1\tilde\delta^{(1)}_2\rangle\langle\tilde\delta^{(1)}_3\tilde\delta^{(1)}_4\rangle\nonumber\\
&&+\langle\tilde\delta^{(1)}_1\tilde\delta^{(1)}_3\rangle\langle\tilde\delta^{(1)}_2\tilde\delta^{(1)}_4\rangle\nonumber\\
&&+\langle\tilde\delta^{(1)}_1\tilde\delta^{(1)}_4\rangle\langle\tilde\delta^{(1)}_3\tilde\delta^{(1)}_2\rangle
.\nonumber
\end{eqnarray}
From this, we see that the right hand side of equation (\ref{fourieravg})
involves $(i+n-1)(i+n-3)\dots(1)$ terms if $(i+n)$ is even.  
If $(i+n)$ is odd,
equation (\ref{fourieravg}) is identically zero.  Finally, we recall
that
\begin{equation}
\label{pspcconvention}
\langle\tilde\delta^{(1)}_1\tilde\delta^{(1)}_2\rangle=
\delta^3_D(\vec{k}_1+\vec{k}_2)P_\delta(k_1)
\end{equation}
Combining these results with equation (\ref{deltatothen}) gives us an
algorithm for evaluating the higher order correlation functions of
the $\delta$ distribution.  We illustrate this by considering the bispectrum.

%%%%%%%%%%%%%%%%%%%%%

\section{The bispectrum}
\label{bispsection}

The bispectrum is the three point equivalent of the power spectrum, i.e. it is
the Fourier transform of the three-point correlation function.  In the
expression
\begin{eqnarray}
\nonumber
\langle\delta(\vec{x}_1)\delta(\vec{x}_2)\delta(\vec{x}_3)\rangle&=&
\langle\int \frac{d^3k_{1\dots 3}}{(2\pi)^{9/2}}
e^{i\sum_{j=1}^3\vec{x}_j\vec{k}_j}
\prod_{l=1}^3\tilde\delta(\vec{k}_l)\rangle\nonumber\\
&=&\int\frac{d^3k_{1\dots 3}}{(2\pi)^{9/2}}
e^{i\sum_{j=1}^3\vec{x}_j\vec{k}_j}
\mathcal{B}(\vec{k}_1,\vec{k}_2,\vec{k}_3)\nonumber
\end{eqnarray}
$\mathcal{B}$ is the bispectrum.  From equations (\ref{deltatothen}) and
(\ref{gaussianodd}), we see that the three-point correlation function will, to
leading order, be made up of three terms like
$$\langle\delta^{(2)}\delta^{(1)}\delta^{(1)}\rangle .$$
We write $\delta^{(2)}$, using equations (\ref{deltasecondorder}),
(\ref{Aeqn}) and (\ref{Beqn}), as
\begin{eqnarray}
\delta^{(2)}&=&\mathcal{D}^a\Big(\partial_i\varphi\partial_i\nabla^2\varphi
+\nabla^2\varphi\nabla^2\varphi\Big)%\nonumber\\
%&&\qquad
+\mathcal{D}^b\Big(\partial_{ij}\varphi\partial_{ij}\varphi
+\nabla^2\varphi\nabla^2\varphi
+2\partial_i\varphi\partial_i\nabla^2\varphi\Big) .\nonumber
\end{eqnarray}
From equation (\ref{fourthfourier}), we see that this means
\begin{eqnarray}
\tilde\delta^{(2)}(\vec{k})&=&
\int d^3k^\prime d^3k^{\prime\prime}
\frac{\tilde\delta(\vec{k}^\prime)\tilde\delta(\vec{k}^{\prime\prime})}{D^2}
\delta^3_D(\vec{k}-\vec{k}^\prime-\vec{k}^{\prime\prime})%\nonumber\\
%&&
\times\Big[\mathcal{D}^a
\Big(1+\frac{\vec{k}^\prime\cdot\vec{k}^{\prime\prime}}{k^{\prime 2}}\Big)%\nonumber\\
%&&\phantom{\times\Big[}
+\mathcal{D}^b\Big(1+2\frac{\vec{k}^\prime\cdot\vec{k}^{\prime\prime}}{k^{\prime 2}}
+\frac{(\vec{k}^\prime\cdot\vec{k}^{\prime\prime})^2}{k^{\prime2}k^{\prime\prime2}}\Big)
\Big]\nonumber\\
&=&\int \frac{d^3k^\prime}{(2\pi)^{3/2}}
\frac{\tilde\delta(\vec{k}^\prime)\tilde\delta(\vec{k}-\vec{k}^\prime)}{D^2}
\Big[\mathcal{D}^a
\Big(1+\frac{\vec{k}^\prime\cdot(\vec{k}-\vec{k}^\prime)}{k^{\prime 2}}\Big)
%\nonumber\\
%&&
+\mathcal{D}^b
\Big(1+2\frac{\vec{k}^\prime\cdot(\vec{k}-\vec{k}^\prime)}{k^{\prime 2}}
%\nonumber\\
%&&\phantom{\times\Big[}\qquad
+\frac{(\vec{k}^\prime\cdot(\vec{k}-\vec{k}^\prime))^2}{k^{\prime 2}(\vec{k}-\vec{k}^\prime)^2}
\Big)\Big]\label{fouriersecondorder}
\end{eqnarray}
We can now average equation (\ref{fouriersecondorder}) with two factors of
$\tilde\delta(\vec{k})$ to find the form of the bispectrum.
%\begin{widetext}
\begin{eqnarray}
\langle\tilde\delta^{(1)}_1\tilde\delta^{(1)}_2\tilde\delta^{(2)}_3\rangle&=&
\int \frac{d^3k_4}{D^2}\langle\tilde\delta^{(1)}_1\tilde\delta^{(1)}_2\tilde\delta^{(1)}_{3-4}
\tilde\delta^{(1)}_4\rangle
\times\Big[\mathcal{D}^a\Big(1
+\frac{\vec{k}_4\cdot(\vec{k}_3-\vec{k}_4)}{k_4^2}\Big)+
\mathcal{D}^b\Big(1+2\frac{\vec{k}_4\cdot(\vec{k}_3-\vec{k}_4)}{k_4^2}
+\frac{(\vec{k}_4\cdot(\vec{k}_3-\vec{k}_4))^2}{k_4^2(\vec{k_3}-\vec{k}_4)^2}
\Big)
\Big]\nonumber\\
&=&\int \frac{d^3k_4}{D^2}
\Big\{P(k_1)P(k_2)\delta^3_{D(1+3-4)}
\delta^3_{D(2+4)}+P(k_1)P(k_2)\delta^3_{D(1+4)}
\delta^3_{D(2+3-4)}+P(k_1)P(k_4)\delta^3_{D(1+2)}
\delta^3_{D(3)}\Big\}
\nonumber\\
&&\times\Big[\mathcal{D}^a\Big(1
+\frac{\vec{k}_4\cdot(\vec{k}_3-\vec{k}_4)}{k_4^2}\Big)
+\mathcal{D}^b\Big(1+2\frac{\vec{k}_4\cdot(\vec{k}_3-\vec{k}_4)}{k_4^2}
+\frac{(\vec{k}_4\cdot(\vec{k}_3-\vec{k}_4))^2}{k_4^2(\vec{k_3}-\vec{k}_4)^2}\Big)
\Big]\label{bispectrumcalc}
\end{eqnarray}
%\end{widetext}
%\begin{eqnarray}
%\langle\tilde\delta^{(1)}_1\tilde\delta^{(1)}_2\tilde\delta^{(2)}_3\rangle&=&
%\int d^3k_4\langle\tilde\delta^{(1)}_1\tilde\delta^{(1)}_2\tilde\delta^{(1)}_{3-4}
%\tilde\delta^{(1)}_4\rangle\nonumber\\
%&&\times\Big[\mathcal{D}^a\Big(1
%+\frac{\vec{k}_4\cdot(\vec{k}_3-\vec{k}_4)}{k_4^2}\Big)+\nonumber\\
%&&\mathcal{D}^b\Big(1+2\frac{\vec{k}_4\cdot(\vec{k}_3-\vec{k}_4)}{k_4^2}
%+\frac{(\vec{k}_4\cdot(\vec{k}_3-\vec{k}_4))^2}{k_4^2(\vec{k_3}-\vec{k}_4)^2}
%\Big)
%\Big]\nonumber\\
%&=&\int d^3k_4
%\Big\{P(k_1)P(k_2)\delta^3_{D(1+3-4)}
%\delta^3_{D(2+4)}\nonumber\\
%&&\phantom{\int}+P(k_1)P(k_2)\delta^3_{D(1+4)}
%\delta^3_{D(2+3-4)}\nonumber\\
%&&\phantom{\int}+P(k_1)P(k_4)\delta^3_{D(1+2)}
%\delta^3_{D(3)}\Big\}
%\nonumber\\
%&&\times\Big[\mathcal{D}^a\Big(1
%+\frac{\vec{k}_4\cdot(\vec{k}_3-\vec{k}_4)}{k_4^2}\Big)
%\nonumber\\
%&&
%+\mathcal{D}^b\Big(1+2\frac{\vec{k}_4\cdot(\vec{k}_3-\vec{k}_4)}{k_4^2}\nonumber\\
%&&\qquad
%+\frac{(\vec{k}_4\cdot(\vec{k}_3-\vec{k}_4))^2}{k_4^2(\vec{k_3}-\vec{k}_4)^2}\Big)
%\Big]\label{bispectrumcalc}
%\end{eqnarray}
(note
$\delta^3_{D(i+j-l)}\equiv\delta^3_D(\vec{k}_i+\vec{k}_j-\vec{k}_l)$)
where we have used equation (\ref{pspcconvention}) to get the power spectra and
$\delta^3_D$ factors.  Note that, by the coefficients of the growth terms,
the $\delta^3_D(\vec{k}_3)$ term will vanish upon integration.  
Once that term is
discarded, we see that the bispectrum is proportional to
$\delta^3_D(\vec{k}_1+\vec{k}_2+\vec{k}_3)$.  Specifically
%\begin{widetext}
\begin{eqnarray}
\mathcal{B}(\vec{k}_1,\vec{k}_2,\vec{k}_3)
&=&\frac{P(k_1)P(k_2)}{D^2}\Big[\mathcal{D}^a(1
+\frac{\vec{k}_2\cdot\vec{k}_1}{k_2^2})+
\mathcal{D}^b(1+2\frac{\vec{k}_2\cdot\vec{k_1}}{k_2^2}
+\frac{(\vec{k}_2\cdot\vec{k}_1)^2}{k_2^2k_1^2}
\Big]\delta^3_D(\vec{k}_1+\vec{k}_2+\vec{k}_3)\nonumber\\
&&+\frac{P(k_1)P(k_2)}{D^2}\Big[\mathcal{D}^a(1
+\frac{\vec{k}_2\cdot\vec{k}_1}{k_1^2})+
\mathcal{D}^b(1+2\frac{\vec{k}_2\cdot\vec{k_1}}{k_1^2}
+\frac{(\vec{k}_2\cdot\vec{k}_1)^2}{k_2^2k_1^2}
\Big]\delta^3_D(\vec{k}_1+\vec{k}_2+\vec{k}_3)\nonumber\\
&&+\text{permutations}.\label{bispectrumeqn}
\end{eqnarray}
%\end{widetext}
It is straightforward to integrate the equations in Section
\ref{eomsectionskew} and determine how $\varpi_0$ affects equation
(\ref{bispectrumeqn}).
It is less straightforward to turn this
calculation into a real-world constraint on $\varpi_0$.

\subsection{Galaxy bias and redshift distortions}
\label{redshiftsection}

It is presently impossible to measure the dark matter density at all points in
space.  
As its name suggests, we cannot see dark matter.  
We can only see the galaxies that form in dark
matter haloes.  Going from astronomical observations to a determination of the
bispectrum (\ref{bispectrumeqn}) requires assumptions about how the galaxy
distribution tracks the dark matter distribution (the ``galaxy bias'') for
which we have little theoretical motivation.  As if that were not hard enough,
we only actually see the galaxies in two dimensions (altitude and azimuth
relative to our telescope).  We infer the radial distance to galaxies by
measuring their redshift and assuming that Hubble's law is valid.  This is a
decent assumption for redshifts of a few.  Unfortunately, it means that
galactic peculiar motions interfere with our determination of galaxies'
positions, giving rise to ``redshift distortions'' in the observed distribution
of galaxies.  It is known how to correct for these effects in calculating
equation (\ref{bispectrumeqn}).  We will do so below.  
Our derivation relies heavily on the $\varpi=0$ calculations presented
by Bernardeau in Section 7 of reference \cite{Bernardeau:2001qr}.

%Unfortunately, we will see that the resulting
%imprecision in $\mathcal{B}$ will be too great to meaningfully compare to the
%data.  Other authors overcome this difficulty through the application of
%Lagrangian perturbation theory, evolving the trajectories of fluid elements
%rather than the perturbative fields $\{\delta,\vec{v},\phi\}$
%\cite{Buchert:1992ya, Buchert:1993xz,
%Bouchet:1992xz,Bouchet:1994xp,Hivon:1994qb,Scoccimarro:2000sp}.  It is not clear that, in the
%absence of a well-defined Poisson equation, we can apply such a method to
%$\varpi\ne 0$ perturbation theory.  Nevertheless, we proceed with the
%corrections in Eulerian perturbation theory (the method used throughout this
%thesis) as an illustration both of the considerations involved and the
%inadequacies of the approach.  

The correction for scale- and time-independent galaxy bias is simple.  
Assume that the excess number
density of galaxies $\delta_g=(n_g-\bar{n}_g)/\bar{n}_g$ relates to the matter
overdensity $\delta$ by
$$\delta_g=\sum_i \frac{b_i\delta^i}{i!}$$
where $b_i$ are (constant) coefficients of the expansion.  If we then write the
overdensity $\delta$ as we did in the discussion leading up to equation
(\ref{deltatothen}), we have, to second order
\begin{equation}
\label{deltag}
\delta_g=b_1\delta^{(1)}+b_1\delta^{(2)}+\frac{b_2}{2}(\delta^{(1)})^2+\dots
\end{equation}
To first order, the expansion is simply $\delta_g=b_1\delta$.

To account for redshift space distortions, we follow Section 7 of reference
\cite{Bernardeau:2001qr} or Section 9.4 of reference \cite{Dodelson:2003ft}.
We denote redshift space (in which radial distance is reckoned from Hubble's
law) by $\vec{x}_s$.  Physical space will remain $\vec{x}$.  Because
observations are made in redshift space, we want to calculate
the bispectrum of the redshift space distribution using the physical space
evolution equations we derived in Section \ref{eomsectionskew}.  We will work
in the plane-parallel approximation in which the sky is flat and the radial
direction is $\hat{z}$.  By Hubble's law, the redshift space $z$ coordinate of
a given galaxy will be
\begin{equation}
\label{hubbleslaw}
z_s=z_x+\frac{v_z}{\mathcal{H}_0}.
\end{equation}
We use the conformal time Hubble parameter because, throughout this work,
$\vec{v}$ has also been calculated as the conformal time velocity.  It is also
true that $\mathcal{H}_0=H_0$ as long as $a_0=1$.  From equation
(\ref{hubbleslaw}), we have
\begin{equation}
\nonumber
\frac{d^3x_s}{d^3x}=\frac{dx_s dy_sdz_s}{dxdydz}
=1+\frac{1}{\mathcal{H}_0}\partial_zv_z .
\end{equation}
Because we do not want our change in coordinates to change the mass of a given
region, we set
\begin{equation}
\label{zconservemass}
(1+\delta_s)d^3x_s=(1+\delta)d^3x
\end{equation}
from which we find
\begin{eqnarray}
\delta_s&=&(1+\delta)\frac{d^3x}{d^3x_s}-1\nonumber\\
&=&\left(1+\delta-\frac{d^3x_s}{d^3x}\right)\frac{d^3x}{d^3x_x}\nonumber\\
&=&\left(\delta-\frac{1}{\mathcal{H}_0}\partial_zv_z\right)\frac{d^3x}{d^3x_s}
.\nonumber
\end{eqnarray}
Now, we can write
\begin{eqnarray}
\tilde\delta_s(\vec{k})&=&\int\frac{d^3x_s}{(2\pi)^{3/2}}
e^{-i\vec{k}\cdot\vec{x}_s}\delta_s(\vec{x}_s)\nonumber\\
&=&\int\frac{d^3x_s}{(2\pi)^{3/2}}
e^{-i\vec{k}\cdot\vec{x}-ik_zv_z/\mathcal{H}_0}
\left(\delta-\frac{1}{\mathcal{H}_0}\partial_zv_z\right)\frac{d^3x}{d^3x_s}
\label{penult}\\
&\approx&\int\frac{d^3x}{(2\pi)^{3/2}}
e^{-i\vec{k}\cdot\vec{x}}\left(1-\frac{ik_zv_z}{\mathcal{H}_0}\right)
\left(\delta-\frac{1}{\mathcal{H}_0}\partial_zv_z\right)
\label{ult}
\end{eqnarray}
where the $\approx$ comes from expanding equation (\ref{penult}) to first order
in perturbed quantities.
If we want the Fourier transform of the galaxy overdensity in redshift
space, equation (\ref{ult}) must be rewritten
%\begin{widetext}
\begin{equation}
\label{redshiftfourier}
\tilde\delta(\vec{k})_{g,s}=
\int\frac{d^3x}{(2\pi)^{3/2}}
e^{-i\vec{k}\cdot\vec{x}}\left(1-\frac{ik_zv_z}{\mathcal{H}_0}\right)
\left(b_1\delta^{(1)}+b_1\delta^{(2)}+\frac{b_2}{2}\delta^{(1)2}
-\frac{1}{\mathcal{H}_0}\partial_zv_z\right).
\end{equation}
%\end{widetext}
To proceed further, we will require expressions for the first and second order
parts of the peculiar velocity
field, $\vec{v}^{(1)}$ and $\vec{v}^{(2)}$, in terms of the Fourier transform
$\tilde\varphi$.

Recalling equations (\ref{vfirstorder}) and (\ref{vsecondorder}), we write
\begin{eqnarray}
\vec{v}^{(1)}&=&-i\dot{D}\int \frac{d^3k}{(2\pi)^{3/2}}e^{i\vec{k}\cdot\vec{x}}
\vec{k}\tilde\varphi(\vec{k})\nonumber\\
\tilde{\vec{v}}^{(1)}&=&-i\dot{D}\vec{k}\tilde\varphi
=i\frac{\dot{D}}{D}\frac{\vec{k}}{k^2}\tilde{\delta}^{(1)}
\label{vfirstfourier}\\
\vec{v}^{(2)}&=&i\int\frac{d^3k_1d^3k_2}{(2\pi)^3}
e^{i\vec{x}\cdot(\vec{k}_1+\vec{k}_2)}\tilde\varphi(\vec{k}_1)\tilde\varphi(\vec{}_2)
%\nonumber\\
%&&
\times\Big[(\dot{\mathcal{D}}^a-D\dot{D})(k_1^2\vec{k_2})
+\dot{\mathcal{D}}^b([\vec{k}_1\cdot\vec{k_2}]\vec{k_2}
+k_1^2\vec{k}_2)\Big]\nonumber\\
\tilde{\vec{v}}^{(2)}&=&i\int\frac{d^3k^\prime}{(2\pi)^{3/2}}\tilde\varphi^\prime
\tilde\varphi(\vec{k}-\vec{k}^\prime)%\nonumber\\
%&&
\times\Big[
(\dot{\mathcal{D}}^a-D\dot{D})(\vec{k}^\prime(\vec{k}-\vec{k}^\prime)^2)
%\nonumber\\
%&&
+\dot{\mathcal{D}}^b(\vec{k}^\prime
[\vec{k}^\prime\cdot(\vec{k}-\vec{k}^\prime)]
+\vec{k}^\prime(\vec{k}-\vec{k}^\prime)^2)\Big].
\label{vsecondfourier}
\end{eqnarray}
Hoping to find a redshift space 
expression equivalent to equation (\ref{bispectrumcalc})
we separate equation (\ref{redshiftfourier}) into
first and second order parts.  For the first order part, we find
\begin{eqnarray}
\tilde{\delta}^{(1)}(\vec{k})_{g,s}&=&b_1\int\frac{d^3xd^3k^\prime}{(2\pi)^3}
e^{i\vec{x}\cdot(\vec{k}^\prime-\vec{k})}(-k^{\prime 2})D
\tilde\varphi(\vec{k}^\prime)%\nonumber\\
%&&\qquad
-\int\frac{d^3xd^3k^\prime}{(2\pi)^3}
e^{i\vec{x}\cdot(\vec{k}^\prime-\vec{k})}\frac{(k^\prime_z)^2}{\mathcal{H}_0}
\dot{D}\tilde\varphi(\vec{k}^\prime)\nonumber\\
&=&(-k^2b_1D-\frac{\dot{D}}{\mathcal{H}_0}k_z^2)\tilde\varphi(\vec{k})
\label{deltagfirstfourier}
\end{eqnarray}
where the factors of $k_z^2$ come from $\partial_zv_z$ (recall that
$\tilde{v}_z\propto k_z$ by equation \ref{vfirstfourier}).

The second order part gives
\begin{eqnarray}
\tilde\delta^{(2)}(\vec{k})_{g,s}&=&b_1\tilde\delta^{(2)}(\vec{k})
%\nonumber\\
%&&
+\frac{b_2}{2}\int\frac{d^3k^\prime}{(2\pi)^{3/2}}D^2
k^{\prime2}(\vec{k}-\vec{k}^\prime)^2\tilde\varphi(\vec{k}^\prime)
\tilde\varphi(\vec{k}-\vec{k}^\prime)\nonumber\\
&&
+\int\frac{d^3k^\prime}{(2\pi)^{3/2}}
\frac{k_z}{\mathcal{H}_0}\tilde\varphi(\vec{k}^\prime)
\tilde\varphi(\vec{k}-\vec{k}^\prime)%\nonumber\\
%&&\qquad
\times\Big[k^\prime_z(\vec{k}-\vec{k}^\prime)^2
(\dot{\mathcal{D}}^a-D\dot{D})%\nonumber\\
%&&\qquad
+\dot{\mathcal{D}}^b
(\vec{k}^\prime\cdot(\vec{k}-\vec{k}^\prime)k^\prime_z
+(\vec{k}-\vec{k}^\prime)^2k^\prime_z)\Big]\nonumber\\
&&
-\frac{\dot{D}}{\mathcal{H}_0}\int\frac{d^3k^\prime}{(2\pi)^{3/2}}
\tilde\varphi(\vec{k}^\prime)\tilde\varphi(\vec{k}-\vec{k}^\prime)
%\nonumber\\
%&&\qquad
\times\Big[-b_1k_zk^\prime_z(\vec{k}-\vec{k}^\prime)^2D
%\nonumber\\
%&&\qquad
-\frac{\dot{D}}{\mathcal{H}_0}k_zk^\prime_z(k_z-k^\prime_z)^2\Big]
\label{deltagsecondfourier}
\end{eqnarray}
where $\tilde\delta^{(2)}(\vec{k})$ on the right hand side of the first line
is given by equation (\ref{fouriersecondorder}).  Combining equations
(\ref{deltagfirstfourier}) and (\ref{deltagsecondfourier}) we can now write, by
analogy with equation (\ref{bispectrumcalc})
%\begin{widetext}
\begin{eqnarray}
\langle\tilde\delta^{(1)}_1\tilde\delta^{(1)}_2\tilde\delta^{(2)}_3\rangle_{g,s}
&=&\int d^3k_4
\Big\{P(k_1)P(k_2)\delta^3_D(\vec{k}_1+\vec{k}_3-\vec{k}_4)
\delta^3_D(\vec{k}_2+\vec{k}_4)\nonumber\\
&&\qquad +P(k_1)P(k_2)\delta^3_D(\vec{k}_1+\vec{k}_4)
\delta^3_D(\vec{k}_2+\vec{k}_3-\vec{k}_4)
+P(k_1)P(k_4)\delta^3_D(\vec{k}_1+\vec{k}_2)
\delta^3_D(\vec{k}_3)\Big\}
\nonumber\\
&&\qquad\times\Big[\frac{b_1}{D^2}\mathcal{D}^a(1
+\frac{\vec{k}_4\cdot(\vec{k}_3-\vec{k}_4)}{k_4^2})+
\frac{b_1}{D^2}\mathcal{D}^b\Big(1+2\frac{\vec{k}_4\cdot(\vec{k}_3-\vec{k}_4)}{k_4^2}
+\frac{(\vec{k}_4\cdot(\vec{k}_3-\vec{k}_4))^2}{k_4^2(\vec{k_3}-\vec{k}_4)^2}
\Big)
\nonumber\\
&&\qquad+\frac{b_2}{2}+\frac{k_{3z}}{D^2\mathcal{H}_0}
\Big(\frac{k_{4z}}{k_4^2}(\dot{\mathcal{D}}^a-D\dot{D})
+\dot{\mathcal{D}}^b
\big(\frac{\vec{k}_4\cdot(\vec{k}_3-\vec{k}_4)k_{4z}}{k_4^2(\vec{k}_3-\vec{k}_4)^2}
+\frac{k_{4z}}{k_4^2}\big)\nonumber\\
&&\qquad-\frac{\dot{D}}{D^2\mathcal{H}_0}
\big(-b_1D\frac{k_{3z}k_{4z}}{k_{4}^2}
-\frac{\dot{D}}{\mathcal{H}_0}
\frac{k_{3z}k_{4z}(k_{3z}-k_{4z})^2}{k_4^2(\vec{k}_3-\vec{k}_4)^2}\big)
\Big]\nonumber\\
&&\qquad\times\left[b_1+\frac{\dot{D}}{D\mathcal{H}_0}\left(\frac{k_{1z}}{k_1^2}\right)\right]
\times\left[b_1+\frac{\dot{D}}{D\mathcal{H}_0}\left(\frac{k_{2z}}{k_2^2}\right)\right]
.\label{bispectrumgfourier}
\end{eqnarray}
%\end{widetext}
The galaxy bispectrum in redshift space $\mathcal{B}_{g,s}$ has the same form
as equation (\ref{bispectrumeqn}) with the appropriate substitution from
equation (\ref{bispectrumgfourier}).  Note that, though the
$b_2\delta^3_D(\vec{k}_3)$ term is not identically zero (as the
$\delta^3_D(\vec{k}_3)$ terms in previous expressions are), 
we are still justified in discarding it, as
we will not be interested in values of $\mathcal{B}_{g,s}$ for which
$\vec{k}_3=0$.

The introduction of redshift distortions into equation
(\ref{bispectrumgfourier}) breaks the isotropy of the $\mathcal{B}$.  Peculiar
velocities
only affect our measurement in the $\hat{z}$ direction, so it matters which
way our $\{\vec{k}_1,\vec{k}_2,\vec{k_3}\}$ triangles are oriented.  We can
average over this orientation-dependence by integrating
%\begin{eqnarray}
%\mathcal{B}_\text{avg}(\vec{k}_1,\vec{k}_2,\vec{k}_3)&=&
%\int_0^\pi
%\sin\theta_1d\theta_1\int_{\theta_1-\theta_{12}}^{\theta_1+\theta_{12}}
%\sin\theta_2d\theta_2
%\mathcal{B}(\vec{k}_1,\vec{k}_2,\vec{k}_3)/N\label{bispavg}\\
%N&=&\int_0^\pi
%\sin\theta_1d\theta_1\int_{\theta_1-\theta_{12}}^{\theta_1+\theta_{12}}
%\sin\theta_2d\theta_2
%\nonumber
%\end{eqnarray}
\begin{equation}
\label{bispavg}
\mathcal{B}_\text{avg}(k_1,k_2,k_3,\theta_{12})=\int_0^\pi
\frac{\sin\theta_1}{4\pi}
d\theta_1\int_0^{2\pi}d\phi \mathcal{B}_{g,s}(\vec{k}_1,\vec{k}_2,\vec{k}_3)
\end{equation}
where $\theta_1$ is the angle that $\vec{k}_1$ makes with the $z$ axis,
$\phi$ is the angle that the plane defined by $\vec{k}_2$ and $\vec{k}_3$ makes
with the plane defined by $\vec{k}_1$ and the $z$ axis, and $\theta_{12}$ is the
angle between $\vec{k}_1$ and $\vec{k}_2$.
Appendix \ref{lptsection} checks the consistency of our expression with prior
results derived in Lagrangian perturbation theory.

%%%%%%%%%%%%%%%%%%%%

\begin{figure}[!t]
\includegraphics[scale=0.3]{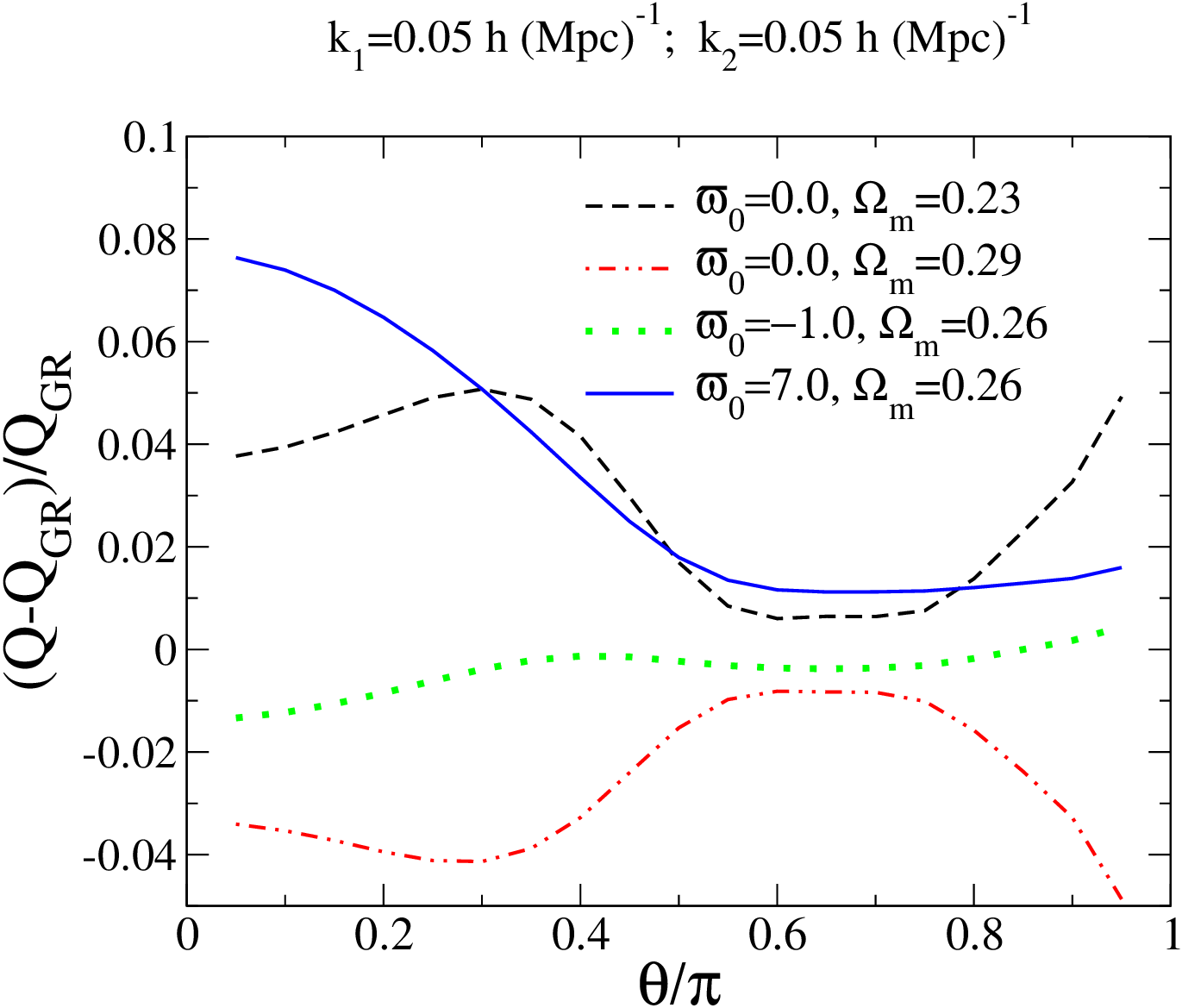}
\caption{
We plot the effect of varying $\varpi_0$ and $\Omega_m$ on the
unbiased normalized bispectrum with redshift distortions, $Q$, 
defined in equation (\ref{qdef}). 
The vertical axis is
the relative deviation from the result in the WMAP 5-year maximum likelihood
GR cosmology.  The horizontal axis is the angle between $\vec{k}_1$ and
$\vec{k}_2$ in units of $\pi$.
Linear power spectra were calculated using the code CMBfast \cite{Seljak:1996is}
modified as in reference \cite{Daniel:2008et}.
We find that $Q$ is much more sensitive to small changes in $\Omega_m$ than it
is to large changes in $\varpi_0$.  This rules out the possibility of
constraining $\varpi_0$ from the three-point correlation function.
}
\label{bispfig}
\end{figure}

To eliminate effects due to initial conditions, it is typical to talk about the
normalized bispectrum $Q$ defined so that
\begin{eqnarray}
Q(\vec{k}_1,\vec{k}_2,\vec{k}_3)&=&
\frac{\mathcal{B}_\text{avg}(k_1,k_2,k_3,\theta_{12})}{b_1^4a_0^2(P_1P_2+P_1P_3+P_2P_3)}
\label{qdef}\\
P_i&\equiv&P_\delta(k_i)\nonumber\\
a_0&\equiv&1+\frac{2}{3}f_1+\frac{1}{5}f_1^2\nonumber\\
f_1&\equiv&\frac{a}{D}\frac{dD}{da} .\nonumber
\end{eqnarray}
Figure \ref{bispfig} plots $(Q-Q_\text{GR})/Q_\text{GR}$ for different
values of $\varpi_0$ and $\Omega_m$.  
$Q_\text{GR}$ is the value of the normalized bispectrum (\ref{qdef})
in a $\varpi=0$ universe with the WMAP 5-year maximum likelihood cosmology
\cite{wmapparams}.
We use a Monte Carlo integrator based on the sobol sequence generator sobseq()
presented in reference \cite{numericalrecipes} to evaluate integral
(\ref{bispavg}). 
The WMAP 5-year team \cite{Dunkley:2008ie} reports $\Omega_m=0.26\pm0.03\,
(1\sigma)$.
Reference \cite{Daniel:2009kr} finds that CMB data alone gives
$\varpi_0 =1.7{}^{+4.0}_{-2.0}\, (2\sigma)$.
We find that the variation in the
bispectrum due to a 1$\sigma$ change in $\Omega_m$ is comparable to
the variation due to a 2$\sigma$ change in $\varpi_0$.
Thus we conclude that
the three-point function will be of little help in constraining the value of
$\varpi_0$.  
Even if future experiments improve our constraint on $\Omega_m$, the fact that
measurements of the galaxy three-point correlation function are only precise
to $\gtrsim10\%$ (see Tables A1-A3
of reference \cite{Kulkarni:2007qu}) means that we are a long way off from being
able to constrain $\varpi_0$ from the distribution of galaxies in the Universe.

\section{Scale dependence}
\label{scalesection}

As was discussed in reference \cite{Daniel:2008et}, 
our scale-independent model of
gravitational slip (\ref{varpieqn}) does not affect the shape of the
power spectrum $P_\delta$.  Any effect on the bispectrum
observed in Figure
\ref{bispfig} must therefore be due
to the renormalization of second order perturbations relative to first
order perturbations in the presence of $\varpi\ne 0$.
Clearly, this effect is weaker than that of actually changing the shape
of $P_\delta$, as varying $\Omega_m$ does.  
Figure \ref{growth2fig} plots the effect of varying $\varpi_0$ on both the
unnormalized second order growth functions $\mathcal{D}^a$ and $\mathcal{D}^b$
(solid lines) and the normalized second order growth functions
$\mathcal{D}^a/D^2$ and $\mathcal{D}^b/D^2$ (dashed lines).  
The normalization greatly reduces
the effect of $\varpi_0$, lending credence to our hypothesis that
the principal effect of scale-independent gravitational slip on the 
bispectrum is through the same renormalization previously 
found for the first order growth
function \cite{Daniel:2009kr}.
Fortunately, the most popular alternative gravity models all predict scale-dependent
effects \cite{Hu:2007pj}.  It is therefore incumbent upon us to consider the effect
of including scale dependence in parametrization (\ref{parametrization}).  In this
section, we find that scale-dependent $\varpi$ does amplify non-GR modifications
to the bispectrum, provided one considers the right combination of
$\{k_1,k_2,k_3\}$.

\subsection{Scale-dependent equations of motion}

Because we want solar system tests to remain consistent with general relativity
\cite{Bertotti:2003rm},
we will work in Fourier space and impose the constraint that
$\varpi\rightarrow0$ as $k\rightarrow\infty$.
In this case, it is an easy matter to calculate the effect of
scale-dependent $\varpi$ on our first order results, 
provided that we rewrite parametrizations
(\ref{parametrization}) and (\ref{varpieqn}) as
\begin{eqnarray}
\tilde\psi&=&(1+\varpi)\tilde\phi\label{scaleparam}\\
\varpi&=&\varpi_0\widetilde{K}(k)a^3\label{scalevarpi}
\end{eqnarray}
with the restriction that $\lim_{k\rightarrow\infty}\widetilde{K}(k)=0$.

To first order,
the equations of motion for $\{\phi,\delta,\vec{v}\}$ remain unchanged under
this reparametrization. 
The only difference in our results is that, under (\ref{scalevarpi})
the growth functions $D$ and $f$ are dependent on the
modulus of the wave vector $\vec{k}$ as well as on redshift.  
To higher orders, the presence of the
non-linear source term (\ref{sourceterm}) forces us to reformulate our approach
towards calculating the bispectrum.

Our ability to write equations of motion (\ref{alphaeqn})--(\ref{curlyD2})
depended on the separability of equation (\ref{delsquaredphi}) into scale- and
redshift-dependent parts.  Using the first order solutions for
$\{\phi,\delta,\vec{v}\}$ we were able to write the second order source term
(\ref{sourceterm}) as
\begin{equation}
\nonumber
S^{(2)}=D(1+\varpi)f\partial_i(\varphi\nabla^2\varphi)
+\dot{D}^2\partial_{ij}(\partial_i\varphi\partial_j\varphi) .
\end{equation}
Under reparametrization (\ref{scaleparam}), we have
\begin{eqnarray}
S^{(2)}&=&\int\frac{d^3k_1d^3k_2}{(2\pi)^3}e^{i\vec{x}(\vec{k}_1+\vec{k}_2)}
\Big\{%\nonumber\\
%&&\phantom{d^3k\Big\{}
D(k_1)(1+\varpi(k_2))f(k_2)
\big(k_1^2\vec{k}_1\cdot\vec{k}_2+k_1^2k_2^2\big)
\nonumber\\
&&\phantom{d^3kd^3ke^{ix(k+k)}\Big\{}
+\dot{D}(k_1)\dot{D}(k_2)
\big[k_1^2k_2^2+(\vec{k}_1\cdot\vec{k}_2)^2%\nonumber\\
%&&\phantom{d^3k\Big\{}
+k_1^2\vec{k_1}\cdot\vec{k}_2
+k_2^2\vec{k_1}\cdot\vec{k_2}\big]\Big\}\tilde\varphi(k_1)\tilde\varphi(k_2) .
\label{sourcescale}
\end{eqnarray}
The integration over wave vectors
(stemming from the fact that reparametrization (\ref{scaleparam}) is defined in
Fourier space)
makes it impossible for us to find a separable
solution for $\{\phi^{(2)},\delta^{(2)},\vec{v}^{(2)}\}$.  This does not, however,
mean that we need be stymied in our attempt to calculate the bispectrum under
(\ref{scaleparam}).

Upon inspection of equation (\ref{bispectrumgfourier}), we see that the bispectrum
accounting for redshift space distortions and galaxy bias is made up of terms of the
following types
\begin{eqnarray}
\langle\tilde\delta^{(2)}\tilde\delta^{(1)}\tilde\delta^{(1)}\rangle\label{d2d1d1}\\
\langle\tilde\delta^{(2)}\tilde\delta^{(1)}\partial_z\tilde{v}^{(1)}_z\rangle\label{d2d1v1}\\
\langle\tilde\delta^{(2)}\partial_z\tilde{v}^{(1)}_z\partial_z\tilde{v}^{(1)}_z\rangle\label{d2v1v1}\\
\langle \partial_z\tilde{v}^{(2)}_z\delta^{(1)}\delta^{(1)}\rangle\label{v2d1d1}\\
\langle \partial_z\tilde{v}^{(2)}_z\delta^{(1)}\partial_z\tilde{v}^{(1)}_z\rangle\label{v2d1v1}\\
\langle \partial_z\tilde{v}^{(2)}_z\partial_z\tilde{v}^{(1)}_z\partial_z\tilde{v}^{(1)}_z\rangle\label{v2v1v1}\\
\langle \delta^{(1)}\delta^{(1)}\delta^{(1)}\delta^{(1)}\rangle\label{d1d1d1d1}\\
\langle \delta^{(1)}\delta^{(1)}\partial_z\tilde{v}^{(1)}_z\partial_z\tilde{v}^{(1)}_z\rangle\label{d1d1v1v1}\\
\langle \partial_z\tilde{v}^{(1)}_z\partial_z\tilde{v}^{(1)}_z
\partial_z\tilde{v}^{(1)}_z\partial_z\tilde{v}^{(1)}_z\rangle .\label{v1v1v1v1}
\end{eqnarray}
As noted above, we can calculate the terms built 
totally out of first order pieces
(\ref{d1d1d1d1},\ref{d1d1v1v1},\ref{v1v1v1v1}) exactly as we did in Section
\ref{redshiftsection}, allowing for the new scale dependence of the relevant growth
functions.  We can calculate the terms with second order parts by modifying the
equations of motion (\ref{delsquaredphi}) and (\ref{deltaddoteqn})
so as to calculate
the ensemble averages $\langle\delta^{(2)}\delta^{(1)}\delta^{(1)}\rangle$, etc.
directly (rather than calculating the second order part, multiplying by the first
order parts, and then taking the ensemble average).  We explain this modification
below.

Consider equation (\ref{deltaddoteqn}).  If we multiply both sides 
by two extra factors of $\delta^{(1)}$
take the Fourier transform and take the ensemble
average, we have
%\begin{widetext}
\begin{equation}
\label{deltaddotscale}
\int d\{\text{Fourier}\}
\Big\{
\langle\ddot{\tilde\delta}^{(2)}_1\tilde\delta^{(1)}_2\tilde\delta^{(1)}_3\rangle
+\mathcal{H}\langle\dot{\tilde\delta}^{(2)}\tilde\delta^{(1)}_2\tilde\delta^{(1)}_3\rangle
\Big\}=
\int d\{\text{Fourier}\}
\Big\{
\langle\tilde{S}^{(2)}_1\tilde\delta^{(1)}_2\tilde\delta^{(1)}_3\rangle
-k_1^2(1+\varpi)\langle\tilde\phi^{(2)}\tilde\delta^{(1)}_2\tilde\delta^{(1)}_3\rangle
\Big\}
\end{equation}
%\end{widetext}
where the integral over $d\{\text{Fourier}\}$ stands for an integral over
$$\frac{d^3xd^3k_{1\dots3}}{(2\pi)^6}e^{i\vec{x}(\vec{k}-\sum_{j=1}^3\vec{k}_j)} $$
which reduces to
\begin{equation}
\label{scalemeasure}
\frac{d^3k_{1\dots3}}{(2\pi)^3}\delta^3_D(\vec{k}-\vec{k_1}-\vec{k_2}-\vec{k_3})
\end{equation}
upon integration over $d^3x$.
Simple differential calculus allows us to write the integrand of
equation (\ref{deltaddotscale})
as a second order differential equation for
$$\langle\tilde\delta^{(2)}\tilde\delta^{(1)}\tilde\delta^{(1)}\rangle$$
which is exactly the form of terms like (\ref{d2d1d1}) in equation
(\ref{bispectrumgfourier}).  For example,
\begin{eqnarray}
\langle\dot{\tilde\delta}_1^{(2)}\tilde\delta^{(1)}_2\tilde\delta^{(1)}_3\rangle
&=&\frac{d}{d\tau}\langle\tilde\delta^{(2)}_1\tilde\delta^{(1)}_2\tilde\delta^{(1)}_3\rangle
-\langle\tilde\delta^{(2)}_1\dot{\tilde\delta}^{(1)}_2\tilde\delta^{(1)}_3\rangle
%\nonumber\\
%&&\phantom{\frac{d}{d\tau}\langle\tilde\delta^{(2)}_1\tilde\delta^{(1)}_2\tilde\delta^{(1)}_3\rangle}
-\langle\tilde\delta^{(2)}_1\tilde\delta^{(1)}_2\dot{\tilde\delta}^{(1)}_3\rangle
\nonumber\\
&=&\frac{d}{d\tau}\langle\tilde\delta^{(2)}_1\tilde\delta^{(1)}_2\tilde\delta^{(1)}_3\rangle
%\nonumber\\
%&&\phantom{\frac{d}{d\tau}\langle}
-(\frac{\dot{D}_2}{D_2}+\frac{\dot{D}_3}{D_3})
\langle\delta^{(2)}_1\tilde\delta^{(1)}_2\tilde\delta^{(1)}_3\rangle
\end{eqnarray}
where we have used $\tilde\delta^{(1)}_i=-k_i^2D(k_i)\tilde\varphi(k_i)$.
We can use equation (\ref{delsquaredphi}) to write a
similar differential equation, which we can evolve to calculate the
$\langle\tilde\phi^{(2)}\tilde\delta^{(1)}\tilde\delta^{(1)}\rangle$
term on the right hand side of equation
(\ref{deltaddotscale}).  As for the source term in (\ref{deltaddotscale}), we can
use equation (\ref{sourcescale}) to write
\begin{eqnarray}
%the S2 in anglebrackets
\langle\tilde{S}^{(2)}\tilde\delta^{(1)}\tilde\delta^{(1)}\rangle&=&
\int\frac{d^3xd^3k_{1\dots 4}}{(2\pi)^{15/2}}
e^{i\vec{x}(\vec{k}-\sum_{j=1}^4\vec{k}_j}\Big\{\dots\Big\}\nonumber\\
&&\qquad\times\langle\tilde\varphi_1\tilde\varphi_2\tilde\delta^{(1)}_3\tilde\delta^{(1)}_4\rangle
\nonumber\\
&=&\int\frac{d^3xd^3k_{1\dots 4}}{(2\pi)^{15/2}}
\delta^3_D(\vec{k}-\sum_{j=1}^4\vec{k}_j)\Big\{\dots\Big\}%\nonumber\\
%&&\qquad
\times\frac{1}{D_1D_2k_1^2k_2^2}P_1P_2%\nonumber\\
%&&\qquad
\times\Big(\delta^3_{D(1+3)}\delta^3_{D(2+4)}%\nonumber\\
%&&\qquad\phantom{\times\Big(}
+\delta^3_{D(1+4)}\delta^3_{D(2+3)}\Big)
\nonumber\\
&=&2\int\frac{d^3k_1d^3k_2}{(2\pi)^{9/2}}\Big\{\dots\Big\}
%\nonumber\\
%&&\qquad
\times\frac{P_1P_2}{D_1D_2k_1^2k_2^2}\delta^3_D(\vec{k})
\label{fouriersourcescale}
\end{eqnarray}
where $\{\dots\}$ represents the combination of $k$ vectors and growth functions in
the integrand of equation (\ref{sourcescale}) symmetrized
in terms of $\vec{k}_1$ and
$\vec{k}_2$.  
The $\delta^3_D(\vec{k})$ in equation
(\ref{fouriersourcescale}) means that the right hand side of equation
(\ref{deltaddotscale}) will be zero unless $\vec{k}=0$, which,
from the form (\ref{scalemeasure}) of the integral measure in equation
(\ref{deltaddotscale}), implies that
the left hand side will be zero unless $\vec{k}_1+\vec{k}_2+\vec{k}_3=0$.
This is simply the familiar result that the bispectrum is defined only for
triangular configurations of $k$ vectors.  Noting that
\begin{equation}
\label{dtov}
\langle\tilde\delta^{(2)}_1\tilde\delta^{(1)}_2\partial_z\tilde{v}^{(1)}_{3z}\rangle
=-\frac{\dot{D}_3}{D_3}\frac{k_{3z}^2}{k_3^2}
\langle\tilde\delta^{(2)}_1\tilde\delta^{(1)}_2\tilde\delta^{(1)}_3\rangle
\end{equation}
we are now able to solve for the terms (\ref{d2d1v1}) and (\ref{d2v1v1}) in
equation (\ref{bispectrumgfourier}).

To calculate the terms in (\ref{bispectrumgfourier}) 
which involve $\partial_z\tilde{v}^{(2)}_z$, we use equation
(\ref{deltadoteqn}) to find
%$$\tilde{v}^{(2)i}=-(\nabla)^{-1}\dot{\tilde\delta}^{(2)}
%-\int\frac{d^3xd^3k_1d^3k_2}{(2\pi)^{9/2}}
%e^{i\vec{x}(\vec{k}-\vec{k}_1-\vec{k}_2)}\tilde\delta^{(1)}_1\tilde{v}^{(1)i}_2 .$$
\begin{equation}
\label{v2eqnscale}
\tilde{\vec{v}}^{(2)}=-(\vec\nabla)^{-1}\dot{\tilde\delta}^{(2)}
-\int\frac{d^3k_1d^3k_2}{(2\pi)^{3/2}}
\delta^3_D(\vec{k}-\vec{k}_1-\vec{k}_2)\tilde\delta^{(1)}_1\tilde{\vec{v}}^{(1)}_2 .
\end{equation}
We can treat the first term on the right hand side of (\ref{v2eqnscale}) the
same way that we treated
$\langle\tilde\delta^{(2)}\tilde\delta^{(1)}\tilde\delta^{(1)}\rangle$ 
terms in equation (\ref{deltaddotscale}).
Note that
\begin{eqnarray}
\langle(\vec{\nabla}^{-1})\dot{\tilde\delta}^{(2)}_1\tilde\delta^{(1)}_1\tilde\delta^{(1)}_3\rangle
&=&\Big[\frac{d}{d\tau}
-\left(\frac{\dot{D}_2}{D_2}+\frac{\dot{D}_3}{D_3}\right)\Big]
\times\langle(\vec{\nabla}^{-1})\tilde\delta^{(2)}_1\tilde\delta^{(1)}_2\tilde\delta^{(1)}_3\rangle .
\label{totaldtauscale}
\end{eqnarray}
The term
$\langle(\vec{\nabla}^{-1})\tilde\delta^{(2)}_1\tilde\delta^{(1)}_2\tilde\delta^{(1)}_3\rangle$
can be found by using the same evolution equations we used to find
$\langle\tilde\delta^{(2)}\tilde\delta^{(1)}\tilde\delta^{(1)}\rangle$
except eliminating the leading factor of $2$ from the source term
(\ref{fouriersourcescale}) and rewriting the $\{\dots\}$ as
\begin{eqnarray}
\Big\{\dots\Big\}&=&D(k_2)f(k_3)(1+\varpi(k_3))k_2^2k_3
+\dot{D}(k_2)\dot{D}(k_3)
(k_2^2k_3+\vec{k}_2\cdot\vec{k}_3k_3).
\label{newscaledependence}
\end{eqnarray}
The factor of $2$ is eliminated because 
$\langle(\vec{\nabla}^{-1})\tilde\delta^{(2)}_1\tilde\delta^{(1)}_2\tilde\delta^{(1)}_3\rangle$
is not symmetric in its wave vector arguments.  We will need to account for this by
summing over all possible arrangements of $\{\vec{k}_1,\vec{k}_2,\vec{k}_3\}$ when
we calculate the bispectrum.  The new scale dependence (\ref{newscaledependence})
is to account for the inverse $\nabla$ operator.  Once the equations of motion have
been integrated, we must also multiply by a factor of $-k_{1z}k_{3z}/k_3$ to account for
the fact that we are interested in the $\partial_z$ 
derivative of the $z$-component 
of $\vec{v}^{(2)}$.

The second term on the right hand side of (\ref{v2eqnscale}) is found (relatively)
simply from the first order solutions for $\vec{v}$ and $\delta$.  Again,
we ultimately want to find a term that looks like
$\langle\partial_z\tilde{v}^{(2)}_{1z}\tilde\delta^{(1)}_2\tilde\delta^{(1)}_3\rangle$.
Our first order solutions give
\begin{eqnarray}
\text{FT}\partial_z\left(\delta^{(1)}_4v^{(1)}_{5z}\right)&=&
i\int\frac{d^3k_4d^3k_5}{(2\pi)^{3/2}}\delta^3_D(\vec{k}-\vec{k}_4-\vec{k}_5)
(k_{4z}+k_{5z})\tilde\delta^{(1)}_4\tilde{v}^{(1)}_{5z}\nonumber\\
&=&
-\int\frac{d^3k_4d^3k_5}{(2\pi)^{3/2}}\delta^3_D(\vec{k}-\vec{k}_4-\vec{k}_5)
(k_{4z}+k_{5z})%\nonumber\\
%&&\qquad\qquad
\frac{\dot{D}_5}{D_5}\frac{k_{5z}}{k_5^2}
\tilde\delta^{(1)}_4\tilde\delta^{(1)}_5 \nonumber
\end{eqnarray}
where FT denotes a Fourier transform.
From here we see that
$\text{FT}\langle\partial_z\left(\delta^{(1)}v^{(1)}\right)\tilde\delta^{(1)}\tilde\delta^{(1)}\rangle$
will contribute terms like
$$-P_2P_3\frac{k_{3z}k_{1z}}{k_3^2}\frac{\dot{D}_3}{D_3}$$
to the bispectrum.  This contribution can be added to the contribution
(\ref{totaldtauscale}) 
to find that, the
$\langle\partial_z\tilde{v}^{(2)}_z\tilde\delta^{(1)}\tilde\delta^{(1)}\rangle$ 
term in the redshift distortion-corrected bispectrum will be
\begin{eqnarray}
\langle\partial_z\tilde{v}^{(2)}_z\tilde\delta^{(1)}\tilde\delta^{(1)}\rangle&=&
\sum\frac{k_{3z}k_{1z}}{k_3}\Big\{%\nonumber\\
%&&
-\left[\frac{d}{d\tau}-\left(\frac{\dot{D}_2}{D_2}+\frac{\dot{D}_3}{D_3}\right)\right]
%\nonumber\\
%&&
\langle(\vec{\nabla}^{-1})\tilde\delta^{(2)}_1\tilde\delta^{(1)}_2\tilde\delta^{(1)}_3\rangle
%\nonumber\\
%&&
+\frac{\dot{D}_3}{k_3D_3}P_2P_3
\Big\}
\end{eqnarray}
where the sum is over the 6 permutations of $\{\vec{k}_1,\vec{k}_2,\vec{k}_3\}$.
This is the term (\ref{v2d1d1}) in equation
(\ref{bispectrumgfourier}).  Using equation (\ref{dtov}), we can also
evaluate terms (\ref{v2d1v1}) and (\ref{v2v1v1}).  Thus, we are able to calculate
the redshift distortion-corrected bispectrum in the case of scale-dependent
gravitational slip.

Because our source term (\ref{sourcescale}) depends on 
the power spectrum at specific length scales
we cannot, as in Section \ref{eomsectionskew}, evolve
our equations of motion (\ref{deltaddotscale}) 
from arbitrarily early times without regards to the shape of
$\{\tilde\phi^{(1)},\tilde\delta^{(1)}\}$ and then add the scale-dependence later.
We must, instead, begin with realistic initial conditions for the perturbation modes.
Since the form of (\ref{scalevarpi}) is such that $\lim_{a\rightarrow 0}\varpi=0$,
we take these initial conditions from the Boltzmann code CMBfast
\cite{Seljak:1996is} with the appropriate background parameters and modified it to
output
$\{\tilde\phi^{(1)},\dot{\tilde\phi}^{(1)},\tilde\delta^{(1)},\dot{\tilde\delta}^{(1)}\}$
at $a=0.1$ (chosen as an epoch early enough that $\varpi\sim0$ and late enough
that quasi-linear structure growth can be expected to set in).  We still assume
that
$\tilde\phi^{(2)}=\dot{\tilde\phi}^{(2)}=\tilde\delta^{(2)}=\dot{\tilde\delta}^{(2)}=0$
at this initial epoch.  With these initial conditions, we can calculate the
redshift distortion-corrected bispectrum (\ref{bispectrumgfourier}) in the case of
scale-dependent $\varpi$.  All that remains is to explore the effect of
parametrization (\ref{scaleparam}) on obsevable statisitics.

\subsection{Effects on observables}

\begin{figure}[!h]
\subfigure[]{
\includegraphics[scale=0.3]{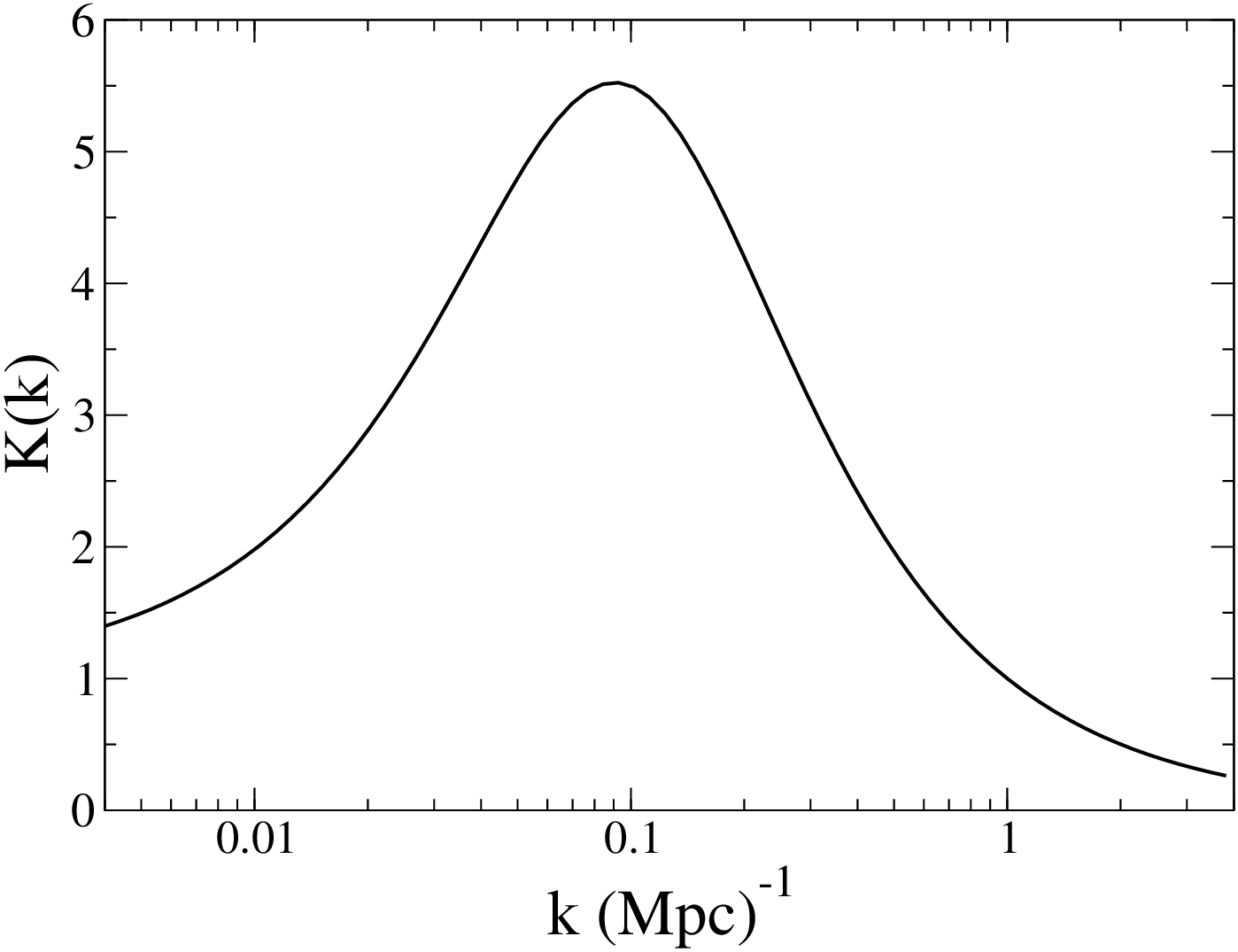}
\label{fofkfig}
}
\subfigure[]{
\includegraphics[scale=0.3]{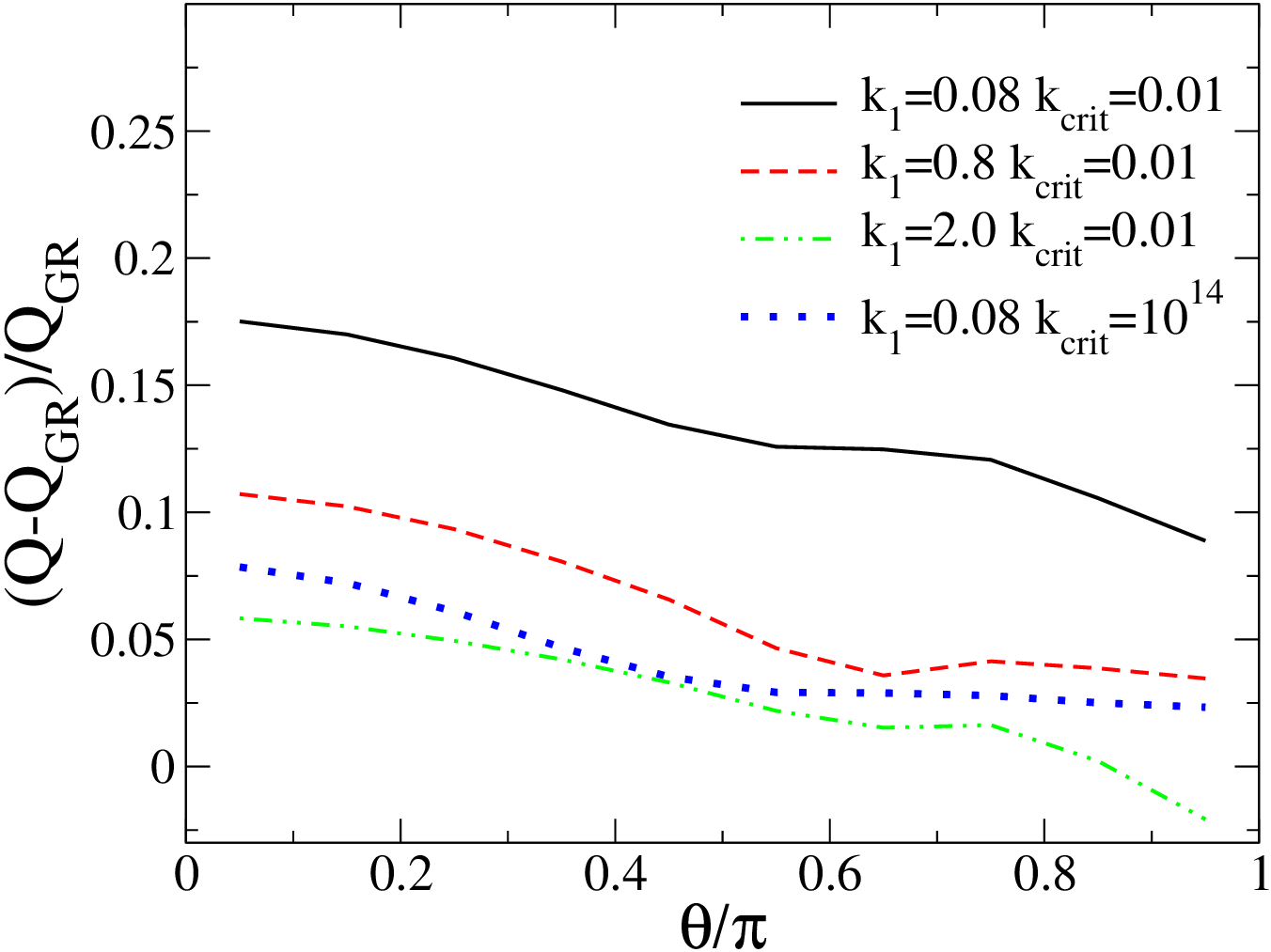}
\label{qscalefig}
}
\caption{The same as Figure \ref{bispfig} except for scale-dependent $\varpi$ of
the form (\ref{scalevarpi}) with $\varpi_0=5$ and 
$\widetilde{K}(k)$ given by equation (\ref{fofkeqn}).
All length scales are in units of Mpc.
All triangles considered are isosceles (i.e., $k_2=k_1$).
As in Figure \ref{bispfig},
$\theta$ is the angle between $\vec{k}_1$ and $\vec{k}_2$.
Figure \ref{fofkfig} plots $\widetilde{K}(k)$.  Figure \ref{qscalefig} plots the
change in the bispectrum relative to the WMAP5 maximum likelihood GR cosmology
\cite{wmapparams} for different triangles.  The case of $k_\text{crit}=10^{14}$ is
shown to illustrate what the curve would look like for scale-independent $\varpi$
(\ref{varpieqn}).  The effect of $\varpi$ is much more pronounced for triangles
whose sides correspond to length scales at which $\widetilde{K}(k)$ peaks.
}
\label{qscalefigtotal}%
\end{figure}

To compare the effects of scale-dependent gravitational slip to a $\varpi=0$  universe,
we perform the same average over orientation as in equation (\ref{bispavg})
and use the same $Q$ statistic as defined in equation (\ref{qdef}).  
The denominator of $Q$
in the case of scale-dependent gravitational slip is written
$$b_1^4(P_1P_2a_1a_2+P_1P_3a_1a_3+P_2P_3a_2a_3)$$
with
\begin{eqnarray}
a_i&\equiv&1+\frac{2}{3}f_{1,i}+\frac{1}{5}f_{1,i}^2\nonumber\\
f_{1,i}&\equiv&\frac{a}{D(k_i)}\frac{dD(k_i)}{da} .\nonumber
\end{eqnarray}
Figure \ref{qscalefigtotal} plots the effect of scale-dependent $\varpi$ on the
$Q$ statistic.  For the purposes of this plot, we choose $\varpi$ to have a scale
dependence given by (recall equation \ref{scalevarpi})
\begin{equation}
\label{fofkeqn}
\widetilde{K}(k)=\frac{1+\frac{k}{k_\text{crit}}}{1+0.01\left(\frac{k}{k_\text{crit}}\right)^2}.
\end{equation}
In Figure \ref{qscalefigtotal} we choose $k_\text{crit}=0.01 (\text{Mpc})^{-1}$
and $\varpi_0=5$.  This is an illustrative model only and is not meant to
represent any specific theory of gravity.
Including the scale dependence (\ref{fofkeqn}) amplifies the effect of
gravitational slip on $Q$, but only for configurations
$\{\vec{k}_1,\vec{k}_2,\vec{k}_3\}$ in which the length scales correspond to the
peak of $\widetilde{K}(k)$.  It therefore seems reasonable to conclude that, while the
bispectrum may be able to tell us something about scale-dependent gravitational
slip, it will only do so if we choose to examine a particularly sensitive range of
$k$.

%%%%%%%%%%%%%%%%
\section{Conclusion}
\label{conclusion}

Contrary to expectations, we have found that quasi-linear order structure growth is
ineffective at constraining scale-independent gravitational slip (\ref{varpieqn}).
However, with judicious triangle choice, we may be able to say something about a
scale-dependent gravitational slip (\ref{scalevarpi}) by measuring the bispectrum.
To date, there has been no attempt to constrain scale-dependent gravitational slip
with even linear-order cosmological data.  Conventional wisdom has been that the data
is still too imprecise to say anything meaningful about
so detailed an effect as scale-dependence.
The next generation of cosmological experiments will hopefully remedy this oversight
\cite{Daniel:2009kr}.
Given the prevalence of scale-dependent effects in alternative gravity theories
\cite{Hu:2007pj}, it may behoove us to find a useful, model-independent
parametrization of the scale dependence $\widetilde{K}(k)$ (\ref{scalevarpi})
and subject these future datasets to the same analysis given in reference
\cite{Daniel:2009kr}.  Depending on what such investigations say, 
the results of Section
\ref{scalesection} may be able to extend our knowledge by bringing to bear
measurements of the three-point correlation function 
on the question of gravitational
slip.

\acknowledgments
The author would like to thank Robert Caldwell for many useful discussions and criticisms.  This work was supported by funding from Dartmouth College's Gordon F. Hull Fellowship.

\vfill

%%%%%%%%%%%%%%%%%%%%%%%%%%%%%%%%%%%%%%%%%%%%%%%%%%%%%%%%%%%%%%%%%%%%%%%%%%%%%%%%%%%%%%%%%%%%
\bibliographystyle{physrev.bst}
\bibliography{bibliography}

%%%%%%%%%%%%%%%%%%%%%%%%%%%%%%%%%%%%%%%%%%%%%%%%%%%%%%%%%%%%%%%%%%%%%%%%%%%%%%%%%%%%%%%%%%%%  

\begin{appendix}

\section{Comparison to Lagrangian perturbation theory}
\label{lptsection}

Equation (\ref{bispectrumgfourier}) was derived using Eulerian perturbation
theory (i.e., we evolved the perturbed fields $\{\phi,\delta,\vec{v}\}$
directly).
Hivon {\it et al}. \cite{Hivon:1994qb} perform the same calculation
using the Lagrangian perturbation theory formalism laid out in references
\cite{Bouchet:1992xz, Bouchet:1994xp,
Buchert:1992ya, Buchert:1993xz}.  Rather than evolving
the perturbed density and velocity fields, Lagrangian perturbation theory
evolves the displacement field 
$\vec{\Psi}(\tau,\vec{q})$ defined as the displacement of a fluid element whose
initial position in space was $\vec{q}$.  Put another way, at some time $\tau$,
the position of a fluid element originating at spatial position $\vec{q}$ and
initial time $\tau_i$ is
\begin{equation}
\label{psidef}
\vec{x}=\vec{q}+\vec{\Psi}(\tau-\tau_i,\vec{q}) .
\end{equation}
From this, we find that the perturbed velocity field of Section
\ref{eomsectionskew} is just
\begin{eqnarray}
\vec{v}&=&\frac{d\vec{x}}{d\tau}=\dot{\vec{\Psi}}(\tau,\vec{q})
+(\vec{v}\cdot\vec\nabla)\vec\Psi(\tau,\vec{q})\nonumber\\
&=&\dot{\vec\Psi}^{(1)}+(\dot{\vec\Psi}^{(1)}\cdot\vec\nabla)\vec\Psi^{(1)}
+\dot{\vec\Psi}^{(2)}+\mathcal{O}(3)\label{vdefLPTnaive}
\end{eqnarray}
where we have set $\tau_i\equiv0$.  
Note that in our notation (as in Section \ref{eomsectionskew}), an overdot
denotes partial differentiation with respect to conformal time $\tau$.  
In Hivon {\it et al}'s
notation, an overdot denotes the Lagrangian derivative
$(\partial_\tau+\vec{v}\cdot\vec\nabla)$.
The same conservation of mass
considerations that prompted us to write equation (\ref{zconservemass}), now
imply
$$\bar{\rho}(1+\delta)d^3x=\bar\rho d^3q.$$
There is no $\delta$ on the right hand side because we assume that the fluid
starts from a homogeneous state.  Algebra and equation (\ref{psidef}) give
%\begin{widetext}
\begin{eqnarray}
\delta&=&|\frac{d^3 q}{d^3x}|-1=
\bigg(1+\vec\nabla\cdot\vec\Psi^{(1)}+\frac{1}{2}
\Big[(\vec\nabla\cdot\vec\Psi^{(1)})^2
-\partial_i\Psi^{(1)}_j\partial_j\Psi^{(1)}_i\Big]+\vec\nabla\cdot\vec\Psi^{(2)}
+\mathcal{O}(3)\bigg)^{-1}
\nonumber\\
&=&-\vec\nabla\cdot\vec{\Psi}^{(1)}+(\vec\nabla\cdot\vec{\Psi}^{(1)})^2
-\vec\nabla\cdot\vec\Psi^{(2)}
-\frac{1}{2}\Big[(\vec\nabla\cdot\vec\Psi^{(1)})^2
-\partial_i\Psi^{(1)}_j\partial_j\Psi^{(1)}_i\Big]
+\mathcal{O}(3)\label{deltaLPTnaive}
\end{eqnarray}
%\end{widetext}
where we have expanded $\Psi$ in the same way that we expanded $\phi$, $\delta$
and $\vec{v}$ in Section \ref{eomsectionskew}.  With equations
(\ref{vdefLPTnaive}) and (\ref{deltaLPTnaive}), we can find the same physical
quantities in Lagrangian perturbation theory (LPT) as we calculated directly in
Eulerian perturbation theory (EPT) in Section \ref{eomsectionskew}.  
The advantage is that, in LPT, the transition
to redshift space is merely a coordinate transformation.
In this section, we show that our EPT results are consistent with LPT results to
second order.
In reference \cite{Bernardeau:2001qr}, Bernardeau claims that this is not the 
case, that LPT and EPT do not agree and that LPT is more consistent with the
results of N-body simulations (see the discussion following his Figure 48).  
This is merely a confusion of terminology.  What Bernardeau meant to claim is
that second order LPT and EPT results differ substantially from third order LPT
results (the difference between ``tree-level'' and ``one-loop level''
perturbation theory in reference \cite{Scoccimarro:1996jy}) 
and that N-body simulations are more consistent with the third order
results.  In Section \ref{bispsection}, we found 
that the effect of $\varpi_0$ on the second order
results is not strong enough to justify extending our calculations out to higher
order.

Before we begin, there is a
subtlety which bears discussion.  All of the $\vec{\Psi}$ functions and their
spatial derivatives in equations (\ref{vdefLPTnaive}) and
(\ref{deltaLPTnaive}) are expressed in terms of $\vec{q}$ coordinates, that is,
in terms of the spatial positions of fluid elements 
at initial time $\tau=0$.  When
we go to derive field equations for $\phi$ and $\vec{\Psi}$, we will be
expressing $\phi$ and its spatial derivatives in terms of $\vec{x}$
coordinates (spatial positions now).  
Assume that $\Psi_j\ll q_j$.
Because we will only be interested in quantities up to second
order, the transformation between $\vec{q}$ and $\vec{x}$
coordinates is a straightforward
Taylor expansion.
If we use $\vec{\bar\Psi}(\vec{x})$ to represent the displacement field as a
function of the present position $\vec{x}$ (see reference
\cite{Hivon:1994qb}), then
\begin{eqnarray}
\vec{\Psi}(\vec{q})&=&\vec{\bar\Psi}(\vec{x})
+\partial_{x_i}\vec{\bar\Psi}(\vec{x})(\vec{q}-\vec{x})_i
+\mathcal{O}(\vec{\bar\Psi}^3)\nonumber\\
&=&\vec{\bar\Psi}(\vec{x})-\bar\Psi_i\partial_{x_i}\vec{\bar\Psi}(\vec{x})
+\mathcal{O}(\vec{\bar\Psi}^3)\label{psitransform}\\
\partial_q\vec{\Psi}(\vec{q})&=&\frac{\partial x}{\partial q}
\partial_x\Big(\vec{\bar\Psi}(\vec{x})-
\bar\Psi_i\partial_{x_i}\vec{\bar\Psi}(\vec{x})
+\mathcal{O}(\vec{\bar\Psi}^3)\Big)\nonumber\\
&=&\Big(1+\partial_q\vec\Psi\Big)\cdot
\Big(\partial_x\vec{\bar\Psi}
-\partial_x\bar{\Psi}_i\partial_{x_i}\vec{\bar\Psi}
%\nonumber\\
%&&\phantom{\Big(1+\partial_q\vec\Psi\Big)\cdot}
-\bar\Psi_i\partial_x\partial_{x_i}\vec{\bar\Psi}
+\mathcal{O}(\vec{\bar\Psi}^3)\Big)
\nonumber\\
&=&\nabla_x\vec{\bar\Psi}-\bar\Psi_i\partial_x\partial_{x_i}\vec{\bar\Psi}
+\mathcal{O}(\vec{\bar\Psi}^3)
\end{eqnarray}
where we have used the zeroth order equation $\vec{q}=\vec{x}$.
Now
\begin{eqnarray}
\delta&=&-\vec\nabla\cdot\vec{\bar\Psi}^{(1)}+(\vec{\bar\Psi}\cdot\vec{\nabla})
\vec{\nabla}\cdot\vec{\bar\Psi}
+(\vec\nabla\cdot\vec{\bar\Psi}^{(1)})^2%\nonumber\\
%&&
-\frac{1}{2}((\vec\nabla\cdot\bar\Psi^{(1)})^2
-\partial_i\bar\Psi^{(1)}_j\partial_j\bar\Psi^{(1)}_i)%\nonumber\\
%&&
-\vec\nabla\cdot\vec{\bar\Psi}^{(2)}+\mathcal{O}(3)\label{deltaLPT}\\
\vec{v}&=&\dot{\vec{\bar\Psi}}^{(1)}
-\dot{\bar\Psi}^{(1)}_i\partial_i\vec{\bar\Psi}^{(1)}
-\bar\Psi_i^{(1)}\partial_i\dot{\vec{\bar\Psi}}^{(1)}%\nonumber\\
%&&
+\dot{\vec{\bar\Psi}}_i^{(1)}\partial_i\vec{\bar\Psi}^{(1)}
+\dot{\vec{\bar\Psi}}^{(2)}
+\mathcal{O}(3) \nonumber\\
&=&\dot{\vec{\bar\Psi}}^{(1)}
-\bar\Psi_i^{(1)}\partial_i\dot{\vec{\bar\Psi}}^{(1)}
+\dot{\vec{\bar\Psi}}^{(2)}
+\mathcal{O}(3) .\label{vdefLPT}
\end{eqnarray}
From here on, we will deal only in $\vec{\bar\Psi}(\vec{x})$ and derivatives
with respect to $\vec{x}$.  Let us now consider the LPT equations of motion.

LPT proceeds from the geodesic equation
$$u^\alpha\nabla_\alpha u^i=0$$
($i$ is a spatial index) and the Poisson equation.  We once again replace the
Poisson equation with the space-time Einstein equation in the form
(\ref{einsteinrewrite1})
$$\nabla^2\left(\dot\phi+\mathcal{H}(1+\varpi)\phi\right)=
\frac{3}{2}\mathcal{H}^2\Omega_m\dot\delta$$
with $\delta$ taken from equation (\ref{deltaLPT}) so that, to first order,
\begin{equation}
\label{einsteinLPT}
\frac{3}{2}\mathcal{H}^2\Omega_m\Big(\vec{\nabla}\cdot\dot{\vec{\bar\Psi}}^{(1)}
\Big)=
-\nabla^2\left(\dot\phi+\mathcal{H}(1+\varpi)\phi\right) .
\end{equation}  
Using metric (\ref{perturbedmetricskew}) and four-velocity
(\ref{fourvelocity}), the geodesic equation taken out to zeroth order in
$1/c^2$ merely returns equation (\ref{vdoteqn})
$$\dot{\vec{v}}+\mathcal{H}\vec{v}+(\vec{v}\cdot\vec{\nabla})\vec{v}
+\vec{\nabla}\phi(1+\varpi)=0 .$$
To first order in $\{\phi,\delta,\vec{v}\}$, this equation gives
\begin{eqnarray}
-\nabla^2\psi&=&\vec\nabla\cdot\dot{\vec{v}}+\mathcal{H}\vec\nabla\cdot\vec{v}\nonumber\\
&=&\vec\nabla\cdot\ddot{\vec{\bar\Psi}}^{(1)}
+\mathcal{H}\vec\nabla\cdot\dot{\vec{\bar\Psi}}^{(1)}
\label{vdoteqnLPT}
\end{eqnarray} 
where we have used equation (\ref{vdefLPT}).
The time derivative of equation (\ref{einsteinLPT}) gives
%\begin{widetext}
\begin{eqnarray}
\frac{3}{2}\mathcal{H}^2\Omega_m\Big(\vec{\nabla}\cdot\ddot{\vec{\bar\Psi}}^{(1)}
-\mathcal{H}\vec\nabla\cdot\dot{\vec{\bar\Psi}}^{(1)}\Big)&=&
-\nabla^2\Big(\ddot\phi+\dot{\mathcal{H}}(1+\varpi)\phi+\mathcal{H}
(1+\varpi)\dot\phi
+\mathcal{H}\dot\varpi\phi\Big)\nonumber\\
\frac{3}{2}\mathcal{H}^2\Omega_m\Big(\vec\nabla\cdot\ddot{\vec{\bar\Psi}}^{(1)}
+\mathcal{H}\vec{\nabla}\cdot\dot{\vec{\bar\Psi}}^{(1)}\Big)
-3\mathcal{H}^3\Omega_m\vec\nabla\cdot\dot{\vec{\bar\Psi}}^{(1)}&=&\nonumber\\
\frac{3}{2}\mathcal{H}^2\Omega_m\left(-\nabla^2\phi(1+\varpi)\right)
+2\mathcal{H}\nabla^2\left(\dot\phi+\mathcal{H}(1+\varpi)\phi\right)&=&
\label{phieomLPT}
\end{eqnarray}
%\end{widetext}
where, in the last line, we substitute from equations (\ref{einsteinLPT})
and (\ref{vdoteqnLPT}).  
Equation (\ref{phieomLPT}) can be rearranged to give the
first order EPT equation (\ref{delsquaredphi}).  
Clearly, the first order result
for $\phi$ is identical in either LPT or EPT formalism.  The same is true of
$\delta$ and $\vec{v}$, given that equations (\ref{deltaLPT}) and
(\ref{vdefLPT}) give $\dot\delta=-\vec\nabla\cdot\vec{v}$, 
which is the first order part of the EPT equation of motion (\ref{deltadoteqn}).
To second order, we illustrate the consistency between LPT and EPT by observing
that the second order parts of equations (\ref{deltaLPT}) and (\ref{vdefLPT})
are such that
\begin{eqnarray}
\dot\delta^{(2)}&=&\partial_i\bar\Psi_i^{(1)}\partial_j\dot{\bar\Psi}_j^{(1)}+
\dot{\bar\Psi}_i\partial_{ij}\bar\Psi_j
+\bar\Psi_i^{(1)}\partial_{ij}\dot{\bar\Psi}_j^{(1)}
%\nonumber\\
%&&
+\partial_i\dot{\bar\Psi}^{(1)}_j\partial_j\bar\Psi_i^{(1)}
-\partial_i\dot{\bar\Psi}_i^{(2)}\nonumber\\
&=&-\partial_iv^{(2)}_i
+\partial_i\bar\Psi_i^{(1)}\partial_j\dot{\bar\Psi}^{(1)}_j
+\dot{\bar\Psi}^{(1)}_i\partial_{ij}\bar\Psi^{(1)}_j\nonumber\\
&=&-\partial_iv^{(2)}_i-\delta^{(1)}\partial_iv^{(1)}_i
-v^{(1)}_i\partial_i\delta^{(1)}\label{eulerequiv}
\end{eqnarray}
which is the second order part of equation (\ref{deltadoteqn}).  
Note that we have used
the assumption that the fluid flow is irrotational so that
$\partial_i\bar\Psi_j=\partial_j\bar\Psi_i$.  
Using result (\ref{eulerequiv}) as
motivation, we will now write the growth functions from reference
\cite{Hivon:1994qb} in terms of
our own and show that our second order solutions are, indeed, equivalent to
those of LPT.
%growthfactor discussion

\begin{figure}[!t]
%\subfigure[]{
\includegraphics[scale=0.3]{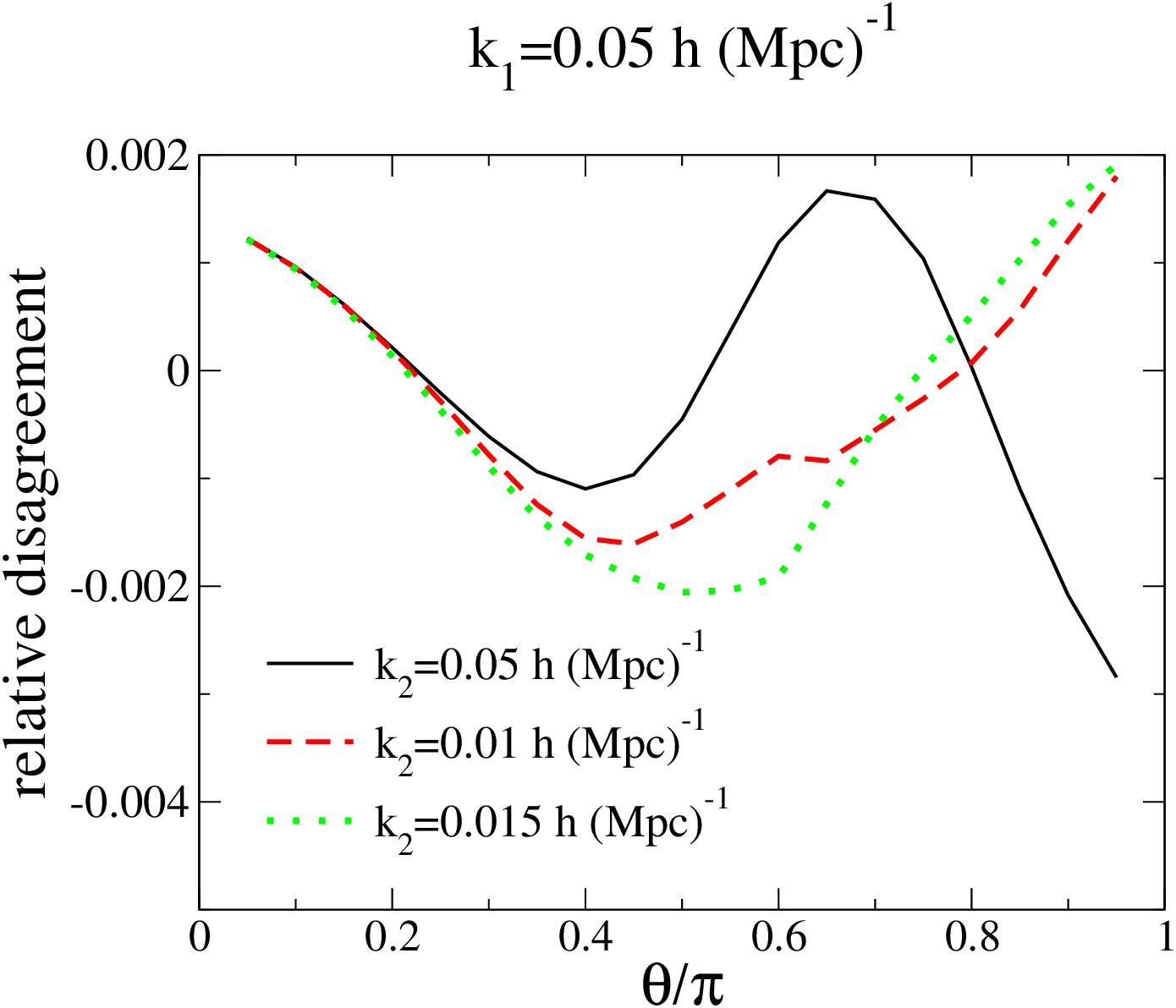}
%\label{myEPTfig}
%}
%\subfigure[]{
%\includegraphics[scale=0.3]{thesisfigs/skewness/lptcomparisonmyscocc.eps}
%\label{myscoccfig}
%}
\caption{We plot the relative disagreement in 
$\mathcal{B}_\text{avg}$ between Eulerian and
Lagrangian (reference \cite{Scoccimarro:2000sp,Hivon:1994qb}) 
perturbation theories.  The horizontal axes are the angle of separation
between the vectors $\vec{k}_1$ and $\vec{k}_2$ in units of $\pi$. 
The vertical axes are
$(\mathcal{B}_\text{avg, EPT}-\mathcal{B}_\text{avg, LPT})/
\mathcal{B}_\text{avg,LPT}$.
%Subfigure \ref{myEPTfig} compares equation
We compare
(\ref{bispavg}) with $b_1=1$ and $b_2=0$ to the results
presented in equations (33)-(35) of reference
\cite{Hivon:1994qb}.
We use equation (\ref{g2eqn}) to express $g_2$ and $f_2=(a/g_2)(dg_2/da)$ in
terms of our growth functions. 
For all curves, $\varpi_0=0$ and the background
cosmology is the WMAP 5-year maximum likelihood universe
\cite{Dunkley:2008ie,wmapparams}.
}
\label{lptcomparisonfig}%
\end{figure}

Results in reference \cite{Hivon:1994qb} are presented 
in terms of the growth factors
$g_1$ and $g_2$ and their dimensionless derivatives $f_i=(a/g_i)(dg_i/da)$. 
These growth factors are defined such that
$$\bar\Psi=g_1\tilde{\bar\Psi}^{(1)}+g_2\tilde{\bar\Psi}^{(2)}$$
where $\tilde{\bar{\Psi}}$ 
denotes the spatial part of $\bar\Psi$.  From the first order
part of equation (\ref{deltaLPTnaive}), it is easy to see that $g_1$ is just our
growth factor $D$ from equation (\ref{deltafirstorder}).  To find $g_2$ as a
function of our second order growth functions $\mathcal{D}^a$ and
$\mathcal{D}^b$, we compare our equation (\ref{fouriersecondorder}) to 
equation (A19) of reference \cite{Hivon:1994qb}, which finds that
$\delta$ goes as
\begin{eqnarray}
\delta^{(2)}&=&g_1^2\int\frac{d^3k_1d^3k_2}{(2\pi)^3}\tilde\phi^{(1)}_1
\tilde\phi^{(1)}_2k_1^2k_2^2
W(|\vec{k}_1+\vec{k}_2|)e^{i\vec{x}\cdot(\vec{k}_1+\vec{k}_2)}%\nonumber\\
%&&\qquad\times
\Big(1+\cos\theta_{12}\frac{k_1}{k_2}
-(1+\frac{g_2}{g_1^2})\frac{(1-\cos^2\theta_{12})}{2}\Big).
\label{delta2hivon}
\end{eqnarray}
The comparison with our equation (\ref{fouriersecondorder}) 
proceeds more directly if we write
\begin{eqnarray}
\mathcal{D}^a&=&-2D_1^2+\frac{3}{2}D_2\nonumber\\
\mathcal{D}^b&=&\frac{3}{2}D_1^2-\frac{3}{4}D_2\nonumber
\end{eqnarray}
where $D_1$ is just the first order growth function $D$ from equation
(\ref{deltafirstorder}), and $D_2$ is defined in equation (39a) of reference
\cite{Bernardeau:1993qu} (this equivalence can be seen by comparing that
work's equation 48 with our equation \ref{fouriersecondorder}).  
Now, we can rewrite our equation as
\begin{eqnarray}
\delta^{(2)}&=&\int\frac{d^3k_1d^3k_2}{(2\pi)^3}\tilde\phi^{(1)}_1
\tilde\phi^{(1)}_2k_1^2k_2^2e^{i\vec{x}\cdot(\vec{k}_1+\vec{k}_2)}\nonumber\\
&&\qquad\times
%&&\phantom{\int d^3k d^3k \phi^{(1)}\phi^{(1)}}
\Big((-2D_1^2+\frac{3}{2}D_2)(1+\frac{k_1}{k_2}\cos\theta_{12})%\nonumber\\
%&&\qquad
+(\frac{3}{2}D_1^2-\frac{3}{4}D_2)
(1+\cos^2\theta_{12}+2\frac{k_1}{k_2}\cos\theta_{12})\Big).%\nonumber\\
\label{delta2rewrite}
\end{eqnarray}
Comparing equations (\ref{delta2hivon}) and (\ref{delta2rewrite}), we find
that
%\begin{eqnarray}
%g_1^2(1-(1+\frac{g_2}{g_1^2})\frac{(1-\cos^2\theta_{12})}{2})&=&
%D_1^2(-\frac{1}{2}+\frac{3}{4}\frac{D_2}{D_1^2}+\frac{3}{2}\cos^2\theta_{12}
%-\frac{3}{4}\frac{D_2}{D_1^2}\cos^2\theta_{12})\nonumber
%\end{eqnarray}
%Simple algebra shows that this equation is satisified if $D_1=g_1$ (as
$D_1=g_1$ (as promised) and
$$g_2=2D_1^2-\frac{3}{2}D_2$$
or
\begin{equation}
\label{g2eqn}
g_2=-3\mathcal{D}^a-4\mathcal{D}^b+2D_1^2
\end{equation}
or
\begin{eqnarray}
\mathcal{D}^{a}&=&-g_2\label{daisg2}\\
\mathcal{D}^{b}&=&\frac{1}{2}(D^2+g_2) .\label{dbisg2}
\end{eqnarray}
Using these results and equation (\ref{vsecondorder}), we find for EPT
\begin{eqnarray}
\vec\nabla\cdot\vec{v}^{(2)}_\text{EPT}&=&(\dot{g}_2+D\dot{D})
(\nabla^2\varphi\nabla^2\varphi+\partial_i\varphi\partial_i\nabla^2\varphi)
%\nonumber\\
%&&
-(\dot{g}_2/2+D\dot{D})
(\nabla^2\varphi\nabla^2\varphi%\nonumber\\
%&&
%\phantom{-(\dot{g}_2/2+D\dot{D})}
+\partial_{ij}\varphi\partial_{ij}\varphi
+2\partial_j\varphi\partial_j\nabla^2\varphi)\nonumber\\
&=&\frac{\dot{g}_2}{2}(\nabla^2\varphi\nabla^2\varphi
-\partial_{ij}\varphi\partial_{ij}\varphi)%\nonumber\\
%&&\qquad
-D\dot{D}(\partial_{ij}\varphi\partial_{ij}\varphi+
\partial_i\varphi\partial_i\nabla^2\varphi) .\nonumber
\end{eqnarray}
For LPT, we note that $\delta^{(1)}=-\partial_i\bar\Psi^{(1)}_i$ means that
$\bar\Psi^{(1)}_i=-D\partial_i\varphi$.  This, combined with the
result from reference \cite{Hivon:1994qb} that 
\begin{equation}
\label{psi2spatial}
\partial_i\bar\Psi^{(2)}_i=\frac{g_2}{g_1^2}\frac{1}{2}
\Big[(\partial_i\bar\Psi_i^{(1)})^2
-\partial_i\bar\Psi_j^{(1)}\partial_j\bar\Psi_i^{(1)}\Big]
\end{equation}
gives
\begin{eqnarray}
\vec\nabla\cdot\vec{v}^{(2)}_\text{LPT}&=&\frac{1}{2}
(\frac{\dot{g}_2}{g_1^2}-2\dot{g}_1\frac{g_2}{g_1^3})
\Big[(\partial_i\bar\Psi^{(1)})^2
-\partial_i\bar\Psi^{(1)}_j\partial_j\bar\Psi^{(1)}_i\Big]%\nonumber\\
%&&
+\dot{g}_1\frac{g_2}{g_1^3}\Big[(\partial_i\bar\Psi^{(1)})^2
-\partial_i\bar\Psi^{(1)}_j\partial_j\bar\Psi^{(1)}_i\Big]%\nonumber\\
%&&
-\partial_j(\bar\Psi^{(1)}_i\partial_i\dot{\bar\Psi}^{(1)}_j)\nonumber\\
&=&\frac{\dot{g}_2}{2}(\nabla^2\varphi\nabla^2\varphi
-\partial_{ij}\varphi\partial_{ij}\varphi)%\nonumber\\
%&&
-D\dot{D}(\partial_{ij}\varphi\partial_{ij}\varphi
+\partial_j\varphi\partial_j\nabla^2\varphi) .\nonumber
\end{eqnarray}
From this we see that equation (\ref{g2eqn}), which guarantees
$\delta^{(2)}_\text{LPT}=\delta^{(2)}_\text{EPT}$ also provides
$\vec{v}^{(2)}_\text{LPT}=\vec{v}^{(2)}_\text{EPT}$.  Figure
\ref{lptcomparisonfig} compares
the results of Section
\ref{bispsection} with the results of LPT in the case of $\varpi=0$.
%end of growthfactor discussion 
LPT results are evaluated from Hivon {\it et al}'s
equations (33)-(35) using our equation (\ref{g2eqn}).
We find that our EPT results agree with Hivon {\it et al}'s
LPT results at the sub-percent level. 
 
\begin{figure}[!t]
\includegraphics[scale=0.3]{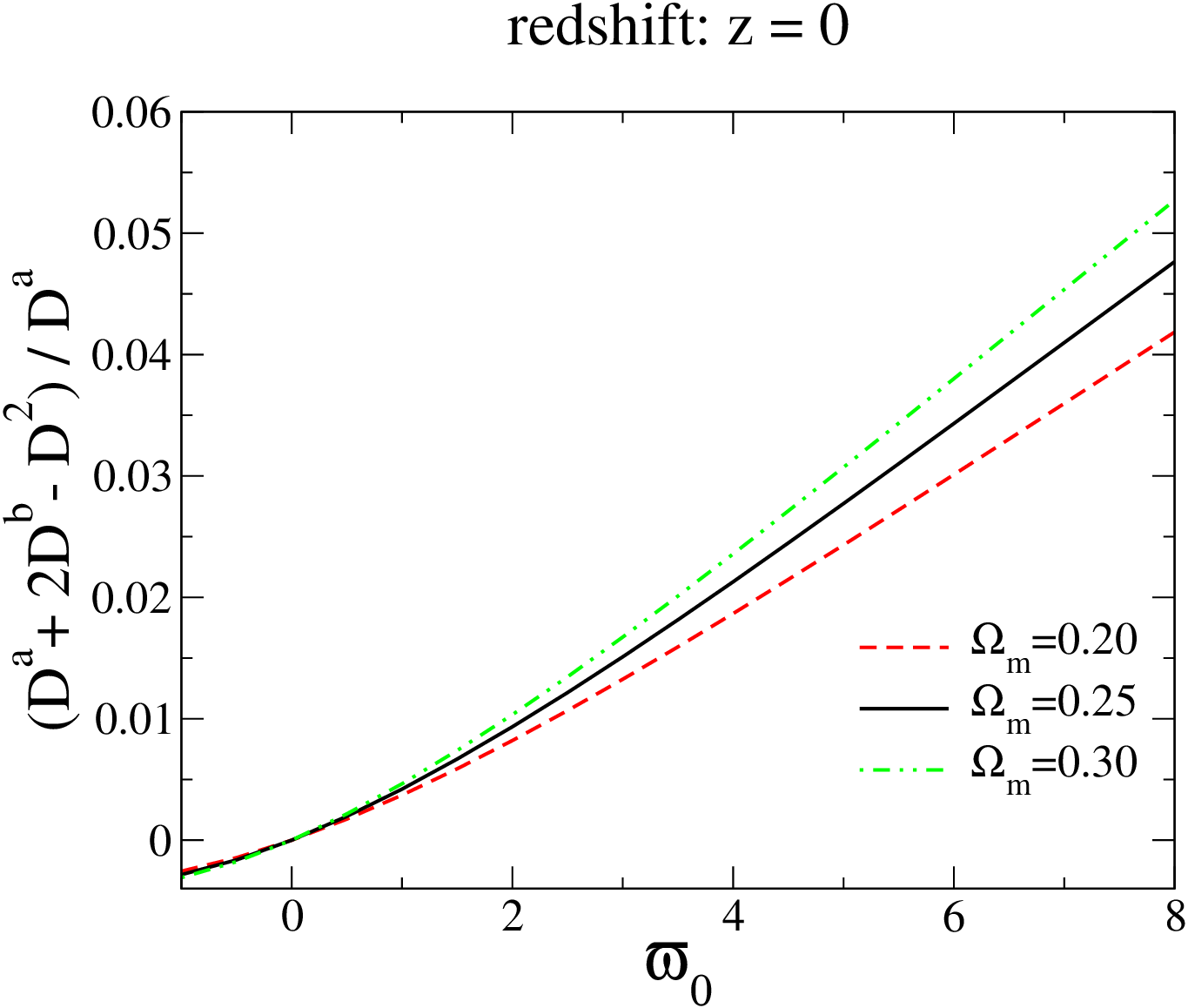}
\caption{
We plot the departure of $\mathcal{D}^a$ from the predictions of equation
(\ref{daisdb}) as a function of $\varpi_0$ for different $\Omega_m$ cosmologies.
While equation (\ref{daisdb}) holds exactly when $\varpi_0=0$, the relationship
breaks down as $|\varpi_0|$ grows.
}
\label{curlyDfig}
\end{figure}

All of the discussion in this Appendix has assumed
$\varpi=0$.  As shown above, our results are perfectly consistent with LPT in
this limit.  One can move between  the two formalisms using the algebraic
relationships (\ref{daisg2}) and (\ref{dbisg2}).
This is not the case once $\varpi_0\ne 0$.  Equations (\ref{daisg2}) and
(\ref{dbisg2}) imply a relationship between $\mathcal{D}^a$, $\mathcal{D}^b$,
and $D$.  Specifically, they imply
\begin{equation}
\label{daisdb}
\mathcal{D}^a=-2\mathcal{D}^b+D^2.
\end{equation}
Figure \ref{curlyDfig} plots the departure from this relationship as a function
of $\varpi_0$ for cosmologies with different values of $\Omega_m$.
All of the curves are evaluated at redshift $z=0$.
While we see
that equation (\ref{daisdb}) is identically true for $\varpi_0=0$, it becomes
less and less true as $|\varpi_0|$ grows.  Recall that equations (\ref{daisg2})
and (\ref{dbisg2}) were derived by assuming that
$\delta^{(2)}_\text{LPT}=\delta^{(2)}_\text{EPT}$ and comparing the growth
functions of the different spatially dependent terms.  If we assume that LPT and
EPT remain in agreement for $\varpi_0\ne0$, it seems
likely that the spatial dependence of $\delta^{(2)}$ must change for
$\varpi_0\ne0$.  Specifically the spatial relationship (\ref{psi2spatial}) may
break down, spoiling the relationship between our EPT growth factors and the LPT
growth factor $g_2$.  Given our success at reproducing the $\varpi_0=0$
solutions of LPT, the main body of this paper analyzed 
the $\varpi_0\ne0$ bispectrum using our EPT
solutions.  It may be worth solving the exact $\varpi_0\ne0$ 
LPT equations of motion in some future work.

%%%%%%%%%%%%%%%%%%%%%%%%%
\section{The skewness}
\label{skewnesssection}

The third order moment (the skewness) is a measure of the
asymmetry of the $\delta$ distribution function once non-linear growth has set
in.  Mathematically, it is defined as
\begin{equation}
S_3=\frac{\langle\delta^3(\vec{x})\rangle}{\langle\delta^2(\vec{x})\rangle^2}\label{skewness}\\
\end{equation}
Kamionkowski and Buchalter derive an exact 
expression for the skewness of the
matter distribution in a general relativistic universe under different
assumptions of flatness and acceleration or deceleration
\cite{Kamionkowski:1998fv}.  In this section, we follow their lead and
derive a similar expression in the case of $\varpi_0\ne0$.

As with the bispectrum, the numerator of the skewness (\ref{skewness}) reduces
to $3\langle\delta^{(2)}(\delta^{(1)})^2\rangle$.  Equation
(\ref{fouriersecondorder}) gives us
%\begin{widetext}
\begin{eqnarray}
\langle\delta^{(2)}(\delta^{(1)})^2\rangle&=&
\int\frac{d^3k_1\dots d^3k_4}{(2\pi)^{6}}
e^{i\vec{x}\cdot(\vec{k}_1+\vec{k}_2+\vec{k}_3+\vec{k}_4)}
\frac{\langle\tilde\delta_1\tilde\delta_2\tilde\delta_3\tilde\delta_4\rangle}{D^2k_3^2k_4^2}\nonumber\\
&&\qquad\times\Bigg[\vec{k}_3\cdot\vec{k_4}k^2_4(\mathcal{D}^a+2\mathcal{D}^b)
+k_3^2
k_4^2(\mathcal{D}^a+\mathcal{D}^b)
+(\vec{k}_3\cdot\vec{k}_4)^2\mathcal{D}^b\Bigg].\label{skewterms}
\end{eqnarray}
Using the following results
\begin{eqnarray}
\int\frac{d^2k_1d^3k_2}{(2\pi)^6}\frac{P_\delta(k_1)P_\delta(k_2)}{D^2}
\Big(-2k_1^2\vec{k}_1\cdot\vec{k}_2-1\Big)&=&
-\int\frac{d^2k_1d^3k_2}{(2\pi)^6}
\frac{P_\delta(k_1)P_\delta(k_2)}{D^2}
\label{negoneterm}\\
\int\frac{d^2k_1d^3k_2}{(2\pi)^6}\frac{P_\delta(k_1)P_\delta(k_2)}{D^2}
\Big(1+2\frac{(\vec{k}_1\cdot\vec{k}_2)^2}{k_1^2k_2^2}\Big)&=&
\int\frac{d^2k_1d^3k_2}{(2\pi)^6}\frac{P_\delta(k_1)P_\delta(k_2)}{D^2}
\Big(1+2\cos^2\theta_{12}\Big)\nonumber\\
\int_{-1}^1d(\cos\theta)\cos^2\theta&=&\frac{2}{3} \nonumber\\
\int\frac{d^2k_1d^3k_2}{(2\pi)^6}\frac{P_\delta(k_1)P_\delta(k_2)}{D^2}
\Big(1+2\cos^2\theta_{12}\Big)&=&
\frac{5}{3}
\int\frac{d^2k_1d^3k_2}{(2\pi)^6}\frac{P_\delta(k_1)P_\delta(k_2)}{D^2}
\label{fivethirdsterm}
\end{eqnarray}
%\end{widetext}
we find that the normalized skewness may be
written
\begin{equation}
\label{KBskewness}
S_3=\frac{3\langle\delta^{(2)}(\delta^{(1)})^2\rangle}
{\langle(\delta^{(1)})^2\rangle^2}=
\frac{3}{D^2}\left(2\mathcal{D}^a+\frac{8}{3}\mathcal{D}^b\right).
\end{equation}
Note that the denominator of the left hand side of equation
(\ref{KBskewness}) is effectively
$$\int\frac{d^3k_1d^3k_2}{(2\pi)^6}P_\delta(k_1)P_\delta(k_2)$$
hence the simplicity of the right hand side.

\begin{figure}[!t]
\subfigure[]{
\includegraphics[scale=0.3]{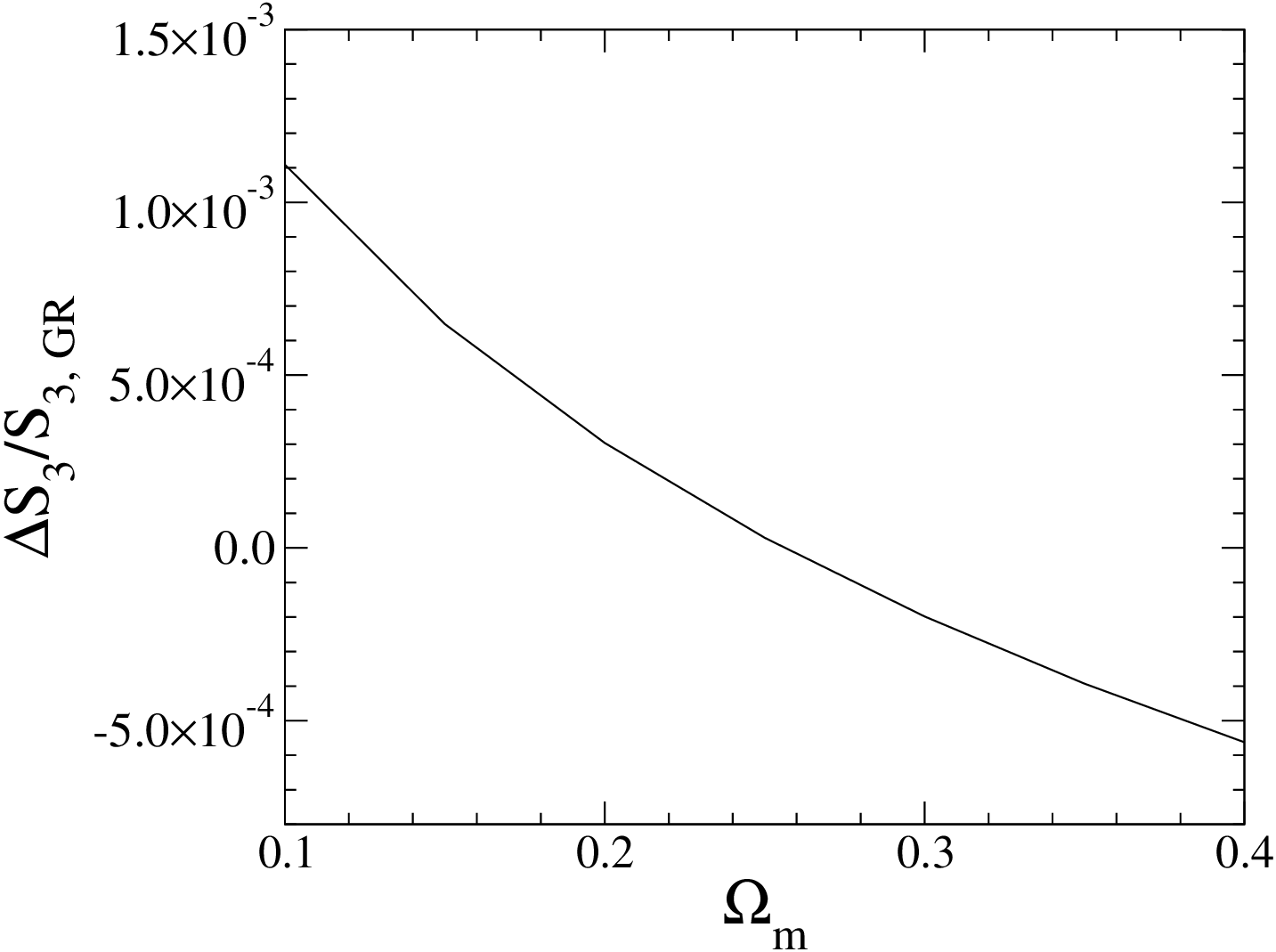}
\label{KBomgfig}
}
\subfigure[]{
\includegraphics[scale=0.3]{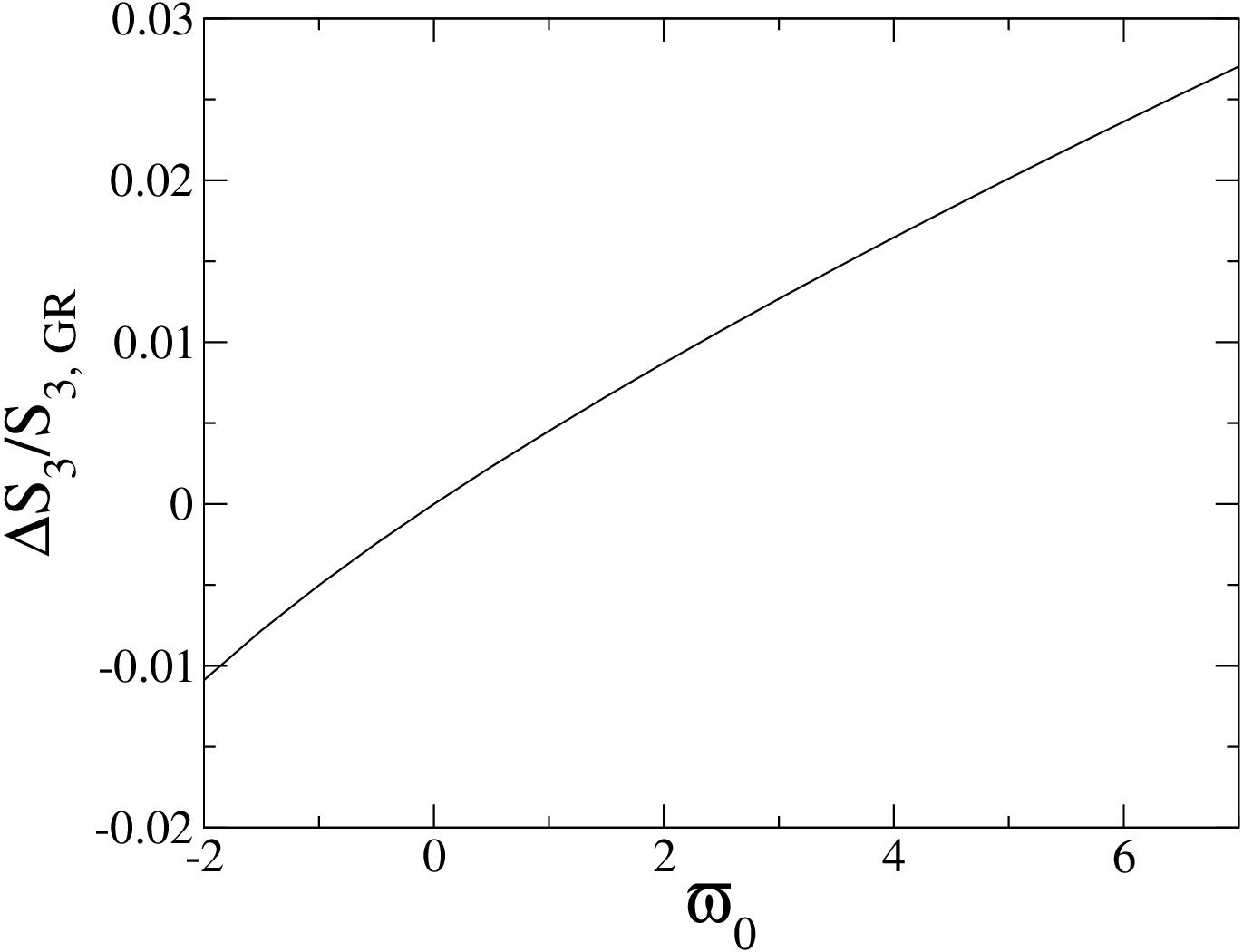}
\label{KBvarpifig}
}
\caption{We plot the change in the skewness $S_3$ (\ref{skewness})
resulting from varying $\Omega_m$ and $\varpi_0$.  When $\Omega_m$ is
varied, $\varpi_0=0$.  When $\varpi_0$ is varied, $\Omega_m=0.256$.
$\Delta S_3$ is assessed relative to a $\varpi_0=0, \Omega_m=0.256$
universe.
All other parameters are 
set to be the WMAP 5-year maximum likelihood values
\cite{Dunkley:2008ie}.
Over regions of
interest, the effect of varying $\varpi_0$ is much stronger than the
effect of varying $\Omega_m$.  Unfortunately, this will not hold when
the calculation is altered to take into account a realistic
observational window function (see Figure \ref{skewfig}).  }
\label{KBfig}%
\end{figure}

Figures \ref{KBomgfig} and \ref{KBvarpifig} plot the effect of
the parameters $\Omega_m$ and $\varpi_0$ on the skewness as
determined by equation (\ref{KBskewness}).  When $\Omega_m$ is
varied, $\varpi_0=0$.  When $\varpi_0$ is varied, $\Omega_m=0.256$.
All other parameters are 
set to be the WMAP 5-year maximum likelihood values
\cite{Dunkley:2008ie}.
As in Figure \ref{bispfig}, the vertical axis is
\begin{equation}
\label{explain}
\frac{\Delta S_3}{S_{3,\text{GR}}}=
\frac{S_3-S_{3|\varpi_0=0,\Omega_m=0.256}}{S_{3|\varpi_0=0,\Omega_m=0.256}} .
\end{equation}
Varying $\varpi_0$ within the WMAP 5-year 2$\sigma$ limit
($\varpi_0=1.4^{+4.0}_{-2.0}\, (2\sigma)$ \cite{Daniel:2009kr})
results in
a few percent variation in $S_3$,
while, for $\varpi_0=0$, variations of $\Omega_m$ within the WMAP 5-year
2$\sigma$ range result in only a few tenths of a percent change in
$S_3$.
The 1$\sigma$ constraint reported by the WMAP 5-year team 
is $\Omega_m=0.26\pm0.03$
\cite{wmapparams}.
Though a few percent is still far too fine a variation to
expect to detect in the data, the fact that the change due to
$\varpi_0$ is so much more pronounced than the change due to $\Omega_m$
leaves open the hope that we may one day be able to distinguish between
an alternative gravity and a dark-energy dominated universe on the
basis of the matter overdensity distribution.  Unfortunately, as we
will see below, even this hope evaporates in the face of considerations
associated with the mechanics of making an observation.

%why is the window function necessary
%why is the KB calculation `unrealistic'

%%%%%%%%%%%%%%%%%%%%%%%%%%%%

\subsection{The smoothed skewness}
\label{skewsmothsection}

As with the bispectrum, we cannot achieve perfect knowledge of the skewness at
any scale.  The birdshot distribution of our mass
gauges (galaxy dynamics, lensing of background galaxies, etc.) forces us to
accept that our measurement of $\delta(\vec{x})$ will have to be averaged over
some scale $R_0$ (e.g. $\sigma_8$ is the rms average mass fluctuation within an
$8h^{-1} \text{Mpc}$ sphere).  
Bernardeau derives expressions for both the skewness and the kurtosis smoothed
over some scale $R_0$ in reference \cite{Bernardeau:1993qu}.
In this section, we will follow his calculation, convolving our results from
Section \ref{skewnesssection} with a top-hat window function.  
We find that whatever hope Figures
\ref{KBomgfig} and \ref{KBvarpifig} give us for constraining $\varpi_0$ vanishes
behind the limitations of actual measurement.

For the purposes of this section, we will work with a simple top-hat window
function $\Theta_{R_0}(r)$ which is zero if $r>R_0$ and unity otherwise.
The smoothed first order overdensity is therefore
\begin{eqnarray}
\delta^{(1)}_{R_0}(\vec{x})&=&\int
d^3x^\prime\delta^{(1)}(\vec{x}^\prime)\Theta_{R_0}(|\vec{x}^\prime-\vec{x}|)\label{dsmothbasic}\\
&=&\int \frac{d^3x^\prime d^3k}{(2\pi)^{3/2}} e^{i\vec{k}\cdot\vec{x}^\prime}\tilde{\delta}^{(1)}(\vec{k})\Theta_{R_0}(|\vec{x}^\prime-\vec{x}|)\nonumber\\
&=&\int \frac{d^3x^\prime d^3k}{(2\pi)^{3/2}} e^{i\vec{k}\cdot(\vec{x}^\prime-\vec{x})}e^{i\vec{k}\cdot\vec{x}}\tilde{\delta}^{(1)}(\vec{k})\Theta_{R_0}(|\vec{x}^\prime-\vec{x}|)\nonumber\\
&=&\int \frac{d^3k}{(2\pi)^{3/2}}\tilde{\delta}^{(1)}(\vec{k})W(kR_0)e^{i\vec{k}\cdot\vec{x}}\label{dsmoth1}\\
W(kR_0)&\equiv&3\left(\frac{\sin(kR_0)}{(kR_0)^3}-\frac{\cos(kR_0)}{(kR_0)^2}\right).\nonumber
\end{eqnarray}
$W(kR_0)$ is the Fourier transform of $\Theta_{R_0}$.  
The second order smoothed overdensity is
\begin{eqnarray}
\nonumber
\delta^{(2)}_{R_0}(\vec{x})&=&\int\frac{d^3k_1d^3k_2}{(2\pi)^3}
\frac{\tilde{\delta}^{(1)}(\vec{k}_1)\tilde{\delta}^{(1)}(\vec{k}_2)}{D^2}
e^{i\vec{x}\cdot(\vec{k}_1+\vec{k}_2)}%\nonumber\\
%&&\qquad\times 
W(R_0|\vec{k}_1+\vec{k}_2|)\nonumber\\
&&\qquad\times
\Bigg[\mathcal{D}^{(a)}\left(1+\frac{\vec{k}_1\cdot\vec{k}_2}{k_1^2}\right)
%&&\qquad\qquad
+\mathcal{D}^{(b)}\left(1+\frac{(\vec{k}_1\cdot\vec{k}_2)^2}{k_1^2k_2^2}
+2\frac{\vec{k}_1\cdot\vec{k}_2}{k_1^2}\right)\Bigg]\label{dsmoth2}
\end{eqnarray}
which is what we would expect from
equation (\ref{fouriersecondorder}) written in a slightly different form
and subjected to the same convolution explicitly shown going from equation
(\ref{dsmothbasic}) to equation (\ref{dsmoth1}).

\begin{figure}[!t]
\includegraphics[scale=0.3]{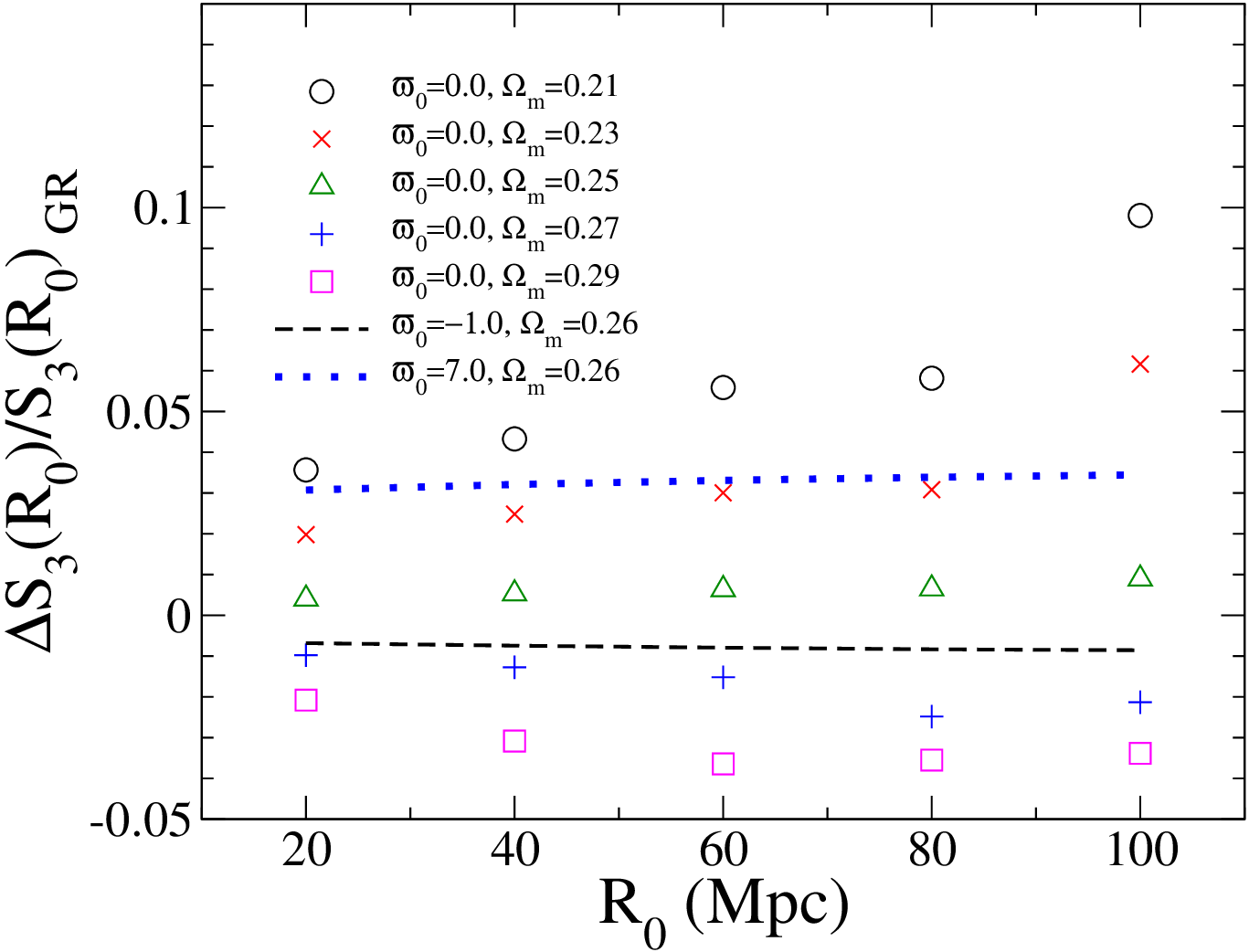}
\caption{We plot the smoothed skewness (\ref{smoothedskewness}) as a function of
scale $R_0$ for different values of $\varpi_0$ and $\Omega_m$.  Results are
presented (as in Figure \protect{\ref{KBfig}}) as a change in skewness relative
to the WMAP 5-year maximum likelihood GR cosmology.  
While $\varpi_0$ retains the strong
influence on $S_3$ observed in Figure \protect{\ref{KBvarpifig}}, the effect of
$\Omega_m$ on the shape of the power spectrum amplifies its influence, so thtat
it is now very difficult to distinguish between a significant departure from
general relativity and a minor shift in $\Omega_m$.  The $1\sigma$ confidence
interval reported by the WMAP 5-year team is $\Omega_m=0.26\pm0.03$.  }
\label{skewfig}%
\end{figure}

By analogy with equation (\ref{skewness}), the smoothed skewness is defined as
\begin{equation}
\label{smoothedskewness}
S_3(R_0)\equiv\frac{\langle\delta_{R_0}(\vec{x})^3\rangle}{\langle\left(\delta^{(1)}_{R_0}(\vec{x})\right)^2\rangle^2}.
\end{equation}
From equations (\ref{dsmoth1}) and (\ref{dsmoth2}), we can write (note that, as
in Bernardeau,
$W_i\equiv W(k_iR_0)$ and $W_{i+j}=W(|\vec{k}_i+\vec{k}_j|R_0)$)
\begin{eqnarray}
3\langle\delta^{(1)2}_{R_0}\delta^{(2)}_{R_0}\rangle&=&
3\int\frac{d^3k_1\dots d^3k_4}{(2\pi)^6}W_1W_2W_{3+4}\frac{1}{D^2}\nonumber\\
&&\phantom{\int 3}\times
\Big\{\mathcal{D}^a\left(1+\frac{\vec{k}_3\cdot\vec{k}_4}{k_3^2}\right)
%\nonumber\\
%&&\phantom{\int 3}
+\mathcal{D}^b\left(1+2\frac{\vec{k}_3\cdot\vec{k}_4}{k_3^2}+\frac{(\vec{k}_3\cdot\vec{k}_4)^2}{k_3^2k_4^2}\right)\Big]
%\nonumber\\
%&&\phantom{\int 3}\times
\langle\tilde\delta_1\tilde\delta_2\tilde\delta_3\tilde\delta_4\rangle.\label{skewnum1}
\end{eqnarray}
Remembering our power spectrum convention (\ref{pspcconvention}), it is useful
to note that the coefficients of both the $\mathcal{D}^a$ and the
$\mathcal{D}^b$ terms will vanish if $\vec{k}_3=-\vec{k}_4$.  Therefore, the
$\langle\tilde\delta_1\tilde\delta_2\rangle\langle\tilde\delta_3\tilde\delta_4\rangle$
term does not contribute to the integral and we have
\begin{eqnarray}
3\langle\delta^{(1)2}_{R_0}\delta^{(2)}_{R_0}\rangle&=&
6\int\frac{d^3k_1\dots d^3k_4}{(2\pi)^6}W_1W_2W_{3+4}\frac{1}{D^2}\nonumber\\
&&\phantom{\int 6}\times\Big\{\mathcal{D}^a\left(1+\frac{\vec{k}_3\cdot\vec{k}_4}{k_3^2}\right)
%\nonumber\\
%&&\phantom{\int 6}
+\mathcal{D}^b\left(1+2\frac{\vec{k}_3\cdot\vec{k}_4}{k_3^2}+\frac{(\vec{k}_3\cdot\vec{k}_4)^2}{k_3^2k_4^2}\right)\Big]\nonumber\\
&&\phantom{\int 6}\times P_\delta(k_1)P_\delta(k_2)\delta^3_D(\vec{k_1}+\vec{k}_3)\delta^3_D(\vec{k}_2+\vec{k}_4)\nonumber\\
&=&6\int\frac{d^3k_1d^3k_2}{(2\pi)^6}W_1 W_2 W_{1+2}\frac{1}{D^2}\nonumber\\
&&\phantom{\int 6}\times\Bigg\{\mathcal{D}^{(a)}\left(1+\frac{\vec k_1\cdot\vec k_2}{k_1^2}\right)
%\nonumber\\
%&&\phantom{\int 6}
+\mathcal{D}^{(b)}\left(1+2\frac{\vec k_1\cdot\vec
k_2}{k_1^2}+\frac{(\vec k_1\cdot\vec k_2)^2}{k_1^2k_2^2}\right)\Bigg\}\nonumber\\
&&\phantom{\int 6}\times P(k_1)P(k_2) .\label{skewnumerator}
\end{eqnarray}
It is tempting to proceed as in Section \ref{skewnesssection} and analytically
evaluate equation (\ref{skewnumerator}) to find the same result as equation
(\ref{KBskewness}) with an added factor of
$$\int\frac{d^3k_1 d^3k_2}{(2\pi)^6}P_\delta(k_1)P_\delta(k_1)W_1W_2W_{1+2}.$$
Unfortunately, the compound window function $W_{1+2}$ depends on the angle
between $\vec{k}_1$ and $\vec{k}_2$, and thus spoils the 
straightforward angular integrations leading to equations (\ref{negoneterm}) and
(\ref{fivethirdsterm}).  Furthermore, the denominator of equation
(\ref{smoothedskewness}) has a completely different dependence on the window
functions and power spectra.  To wit
\begin{equation}
\label{skewdenominator}
\langle(\delta^{(1)}_{R_0})^2\rangle^2
=\left(\int\frac{d^3k}{(2\pi)^{3}}P_\delta(k)W^2(kR_0)\right)^2.
\end{equation}
The smoothed skewness (\ref{smoothedskewness}) cannot be written as a simple
function of growth factors.  It must be integrated over the 6 dimensional
parameter space $\{\vec{k}_1,\vec{k}_2\}$,
which can be reduced to a 4 dimensional parameter space upon noting that
the integrand in equation (\ref{skewnumerator}) does not 
explicitly depend on the azimuthal angles
$\phi_{k_1}$ or $\phi_{k_2}$ so that we can replace those two integrations with
a factor of $4\pi^2$.  
We use the same Monte Carlo integrator we used in Section
\ref{bispsection}.  We present our results in Figure (\ref{skewfig}) as a
change in skewness (relative to a WMAP 5-year maximum likelihood 
GR universe; see
equation \ref{explain}).  While $\varpi_0$ retains the strong
influence on $S_3$ observed in Figure \ref{KBvarpifig}, the effect of
$\Omega_m$ on the shape of the power spectrum amplifies its influence, so that
it is now very difficult to distinguish between a significant departure from
general relativity and a minor shift in $\Omega_m$.
We perform a similar calculation for $S_4(R_0)$ in Section
\ref{kurtosissection}.

%%%%%%%%%%%%%%%%%%%%%%%%%%%%%%%%%%%%%%

\section{Fourth order moments}
\label{kurtosissection}

The smoothed kurtosis $S_4(R_0)$ 
of the overdensity distribution is the fourth order moment
of the distribution averaged over a radius $R_0$, i.e.
\begin{equation}
\label{smoothedkurtosis}
S_4(R_0)=\frac{\langle\delta_{R_0}(\vec{x})^4\rangle-3\langle\delta_{R_0}(\vec{x})^2\rangle^2}{\langle\delta_{R_0}(\vec{x})^2\rangle^3}.
\end{equation}
In this section, we will show the effect of $\varpi\ne0$ on $S_4(R_0)$.
While the effect is stronger than that found on the skewness in Section
\ref{skewnesssection}, this is of little comfort because, as a higher order
moment, the kurtosis is much harder to measure than the (already difficult)
skewness.  Also, we find that the effect of varying $\Omega_m$ is similarly
amplified so that it would still be difficult to discern a large change in
$\varpi_0$ from a small change in $\Omega_m$.

\subsection{Third order equations of motion}
\label{eomappendix}

Calculating the kurtosis requires first that we find the evolution of the third
order perturbations $\{\phi^{(3)},\delta^{(3)},\vec{v}^{(3)}\}$.
We do so in this section.
For easy reference, we recall the first and second order solutions
\begin{eqnarray}
\phi^{(1)}&=&f(\tau)\varphi(\vec x)\nonumber\\
\delta^{(1)}&=&D(\tau)\nabla^2\varphi\nonumber\\
\vec{v}^{(1)}&=&-\dot{D}\vec\nabla\varphi(\vec x)\nonumber\\
\phi^{(2)}&=&\alpha(\tau)A(\vec x)+\beta(\tau)B(\vec x)\nonumber\\
\delta^{(2)}&=&\mathcal{D}^{(a)}(\tau)\nabla^2A+\mathcal{D}^{(b)}(\tau)\nabla^2B\nonumber\\
\vec{v}^{(2)}&=&(-\dot{\mathcal{D}}^a+D\dot{D})\vec{\nabla}A-\dot{\mathcal{D}}^b\vec{\nabla}B\nonumber\\
\nabla^2A&=&\partial_i\left(\nabla^2\varphi\partial_i\varphi\right)\nonumber\\
\nabla^2B&=&\partial_{ij}\left(\partial_i\varphi\partial_j\varphi\right).\nonumber
\end{eqnarray}
The third-order part of equation (\ref{eom}) is
\begin{eqnarray}
\ddot{\delta}^{(3)}+\mathcal{H}\dot{\delta}^{(3)}&=&(1+\varpi)\left(\partial_i\delta^{(2)}\partial_i\phi^{(1)}+\partial_i\delta^{(1)}\partial_i\phi^{(2)}\right)
%\nonumber\\
%&&
+\partial_{ij}\left[2v^{i(2)}v^{j(1)}+\delta^{(1)}v^{i(1)}v^{j(1)}\right]
\nonumber\\
&&
+(1+\varpi)\big(\delta^{(1)}\nabla^2\phi^{(2)}+\delta^{(2)}\nabla^2\phi^{(1)}
%\nonumber\\
%&&\qquad\qquad\qquad
+\nabla^2\phi^{(3)}\big).\label{eomthirdorder}
\end{eqnarray}
Using the recipe outlined in Section \ref{eomsectionskew}, we find
\begin{eqnarray}
\phi^{(3)}&=&\sum_{i=1}^7F^{(i)}(\tau)\mathcal{G}^{(i)}(\vec{x})\nonumber\\
\delta^{(3)}&=&\sum_{i=1}^7\mathcal{F}^{(i)}(\tau)\nabla^2\mathcal{G}^{(i)}(\vec{x})\label{deltathirdorder}
\end{eqnarray}
where
\begin{eqnarray}
\nabla^2\mathcal{G}^{(1)}&=&\partial_i\left[\partial_i\varphi\nabla^2A\right]\nonumber\\
\nabla^2\mathcal{G}^{(2)}&=&\partial_i\left[\partial_i\varphi\nabla^2B\right]\nonumber\\
\nabla^2\mathcal{G}^{(3)}&=&\partial_i\left[\nabla^2\varphi\partial_iA\right]\nonumber\\
\nabla^2\mathcal{G}^{(4)}&=&\partial_i\left[\nabla^2\varphi\partial_iB\right]\label{Gfns}\\
\nabla^2\mathcal{G}^{(5)}&=&\partial_{ij}\left[\partial_j\varphi\partial_iA\right]\nonumber\\
\nabla^2\mathcal{G}^{(6)}&=&\partial_{ij}\left[\partial_j\varphi\partial_iB\right]\nonumber\\
\nabla^2\mathcal{G}^{(7)}&=&\partial_{ij}\left[\nabla^2\varphi\partial_i\varphi\partial_j\varphi\right].\nonumber
\end{eqnarray}
\noindent
The equations of motion for $F^{(i)}$ and $\mathcal{F}^{(i)}$ are
\begin{eqnarray}
\ddot{F}^{(i)}&=&-\dot{F}^{(i)}\mathcal{H}(3+\varpi)%\nonumber\\
%&&
-F^{(i)}\left[\mathcal{H}^23(1+\varpi)(1-\Omega_m)+\mathcal{H}\dot{\varpi}\right]
%\nonumber\\
%&&
+\frac{3}{2}\Omega_m\mathcal{H}^2S^{(3,i)}\nonumber\\
\ddot{\mathcal{F}}^{(i)}&=&-\mathcal{H}\dot{\mathcal{F}}^{(i)}+(1+\varpi)F^{(i)}+S^{(3,i)}.\nonumber
\end{eqnarray}
\noindent
The  right hand side of equation (\ref{eomthirdorder}) gives
\begin{eqnarray}
S^{(3,1)}&=&(1+\varpi)f\mathcal{D}^{(a)}\nonumber\\
S^{(3,2)}&=&(1+\varpi)f\mathcal{D}^{(b)}\nonumber\\
S^{(3,3)}&=&(1+\varpi)D\alpha\nonumber\\
S^{(3,4)}&=&(1+\varpi)D\beta\nonumber\\
S^{(3,5)}&=&-2\dot{D}(-\dot{\mathcal{D}}^a+\dot{D}D)\nonumber\\
S^{(3,6)}&=&2\dot{D}\dot{\mathcal{D}}^b\nonumber\\
S^{(3,7)}&=&D\dot{D}^2 .\nonumber
\end{eqnarray}

%%%%%%%%%%%%%%%%%%%%%%%%%%%%%%%%%
\subsection{The kurtosis}
\label{kurtosissubsection}
The numerator of equation (\ref{smoothedkurtosis}) is equivalent to
\begin{equation}
\nonumber
\langle\delta_{R_0}(\vec{x})^4\rangle-3\langle\delta_{R_0}(\vec{x})^2\rangle^2=
4\langle\delta^{(1)3}\delta^{(3)}\rangle+6\langle
\delta^{(1)2}\delta^{(2)2}\rangle .
\end{equation}
Using equation (\ref{deltathirdorder}) and drawing an analogy with equation (\ref{skewnumerator}),
we find that
\begin{eqnarray}
\langle\delta^{(1)3}\delta^{(3)}\rangle&=&4\int\frac{d^3k_1d^3k_2d^3k_3}{(2\pi)^9}\frac{P(k_1)P(k_2)P(k_3)}{D^3}%\nonumber\\
%&&\qquad\times 
W_1W_2W_3W_{1+2+3}%\nonumber\\
%&&\qquad\times
\sum_{i=1}^{7}\mathcal{F}^{(i)}\sum_{\text{perm}}\tilde{\mathcal{G}}^{(i)}(\vec k_1, \vec k_2, \vec k_3)
\label{d1d3bracket}
\end{eqnarray}
where $\sum_{\text{perm}}\tilde{\mathcal{G}}^{(i)}=\tilde{\mathcal{G}}^{(i)}(\vec k_1, \vec k_2, \vec k_3)+\tilde{\mathcal{G}}^{(i)}(\vec k_1, \vec k_3, \vec k_2)+\dots$ and the $\tilde{\mathcal{G}}^{(i)}$ are the Fourier transforms of equations (\ref{Gfns}) with all factors of $\varphi$ artificially removed.

\begin{figure}[!h]
\includegraphics[scale=0.3]{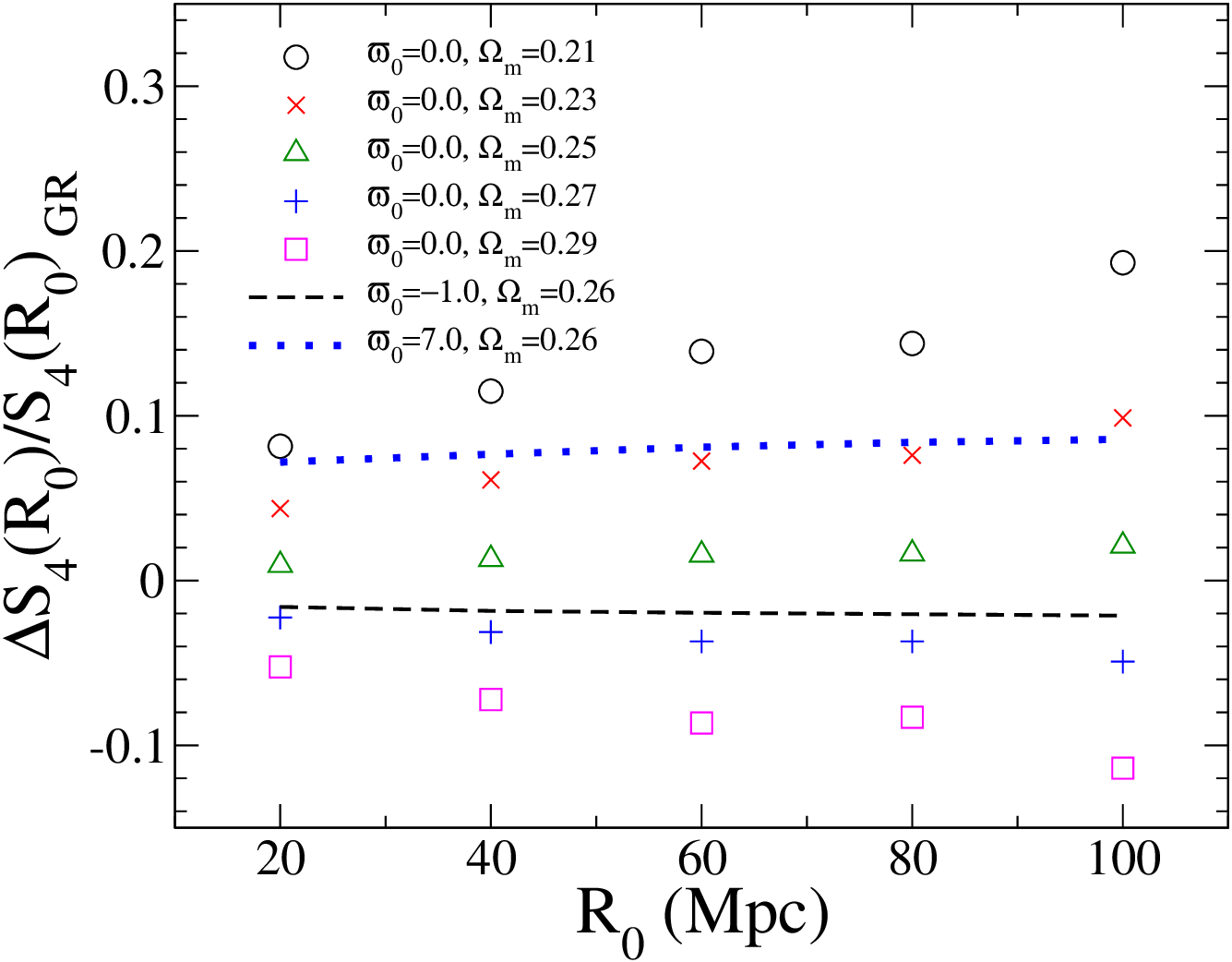}
\caption{We plot the smoothed kurtosis (\ref{smoothedkurtosis}) 
as a function of
scale $R_0$ for different values of $\varpi_0$ and $\Omega_m$.  Results are
presented (as in Figure \protect{\ref{bispfig}}) as a change in kurtosis relative
to the WMAP 5-year maximum likelihood GR cosmology.  The $1\sigma$ confidence
interval reported by the WMAP 5-year team is $\Omega_m=0.26\pm0.03$.}
\label{kurtfig}%
\end{figure}

Similarly, the other part of the kurtosis reduces to
\begin{eqnarray}
\langle\delta^{(1)2}\delta^{(2)2}\rangle&=&6\int\frac{d^3k_1 d^3k_2d^3k_3}{(2\pi)^9}\frac{P(k_1)P(k_2)P(k_3)}{D^4}W_1 W_2%\nonumber\\
%&&\times
\sum_\text{perm}W_{3-1}W_{-3-2}\tilde{\mathcal{D}}(\vec k_3,-\vec k_1)\tilde{\mathcal{D}}(-\vec k_3, -\vec k_2)
\label{d1d2bracket}\\
\tilde{\mathcal{D}}(\vec k_i, \vec k_j)&=&k_i^2k_j^2\Bigg[\mathcal{D}^{(a)}\left(1+\frac{\vec k_i\cdot \vec k_j}{k_i^2}\right)%\nonumber\\
%&&
+\mathcal{D}^{(b)}\left(1+2\frac{\vec k_i\cdot\vec k_j}{k_i^2}
+\frac{(\vec k_i\cdot\vec k_j)^2}{k_i^2k_j^2}\right)\Bigg] .\label{kurtdterms}
\end{eqnarray}
In this case 
$\sum_\text{perm}W_{3-1}W_{-3-2}\tilde{\mathcal{D}}(\vec k_3,-\vec k_1)\tilde{\mathcal{D}}(-\vec k_3, -\vec k_2)$ 
has 8 terms, permuting over both the content of the two pairs of vectors
\begin{equation}
\nonumber
\tilde{\mathcal{D}}(\vec k_3,-\vec k_1)\tilde{\mathcal{D}}(-\vec k_3, -\vec k_2)\rightarrow \tilde{\mathcal{D}}(\vec k_3,-\vec k_2)\tilde{\mathcal{D}}(-\vec k_3, -\vec k_1)\text{, etc.}
\end{equation} 
and permuting over the vectors' orders within the pairs
\begin{equation}
\nonumber
\tilde{\mathcal{D}}(\vec k_3,-\vec k_1)\tilde{\mathcal{D}}(-\vec k_3, -\vec k_2)\rightarrow \tilde{\mathcal{D}}(-\vec k_1,\vec k_3)\tilde{\mathcal{D}}(-\vec k_2, -\vec k_3) \text{,  etc.}
\end{equation} 
Note also that $\vec k_3$ is always positive in the first pair and negative in the second.  
$\vec k_1$ and $\vec k_2$ are always negative.  These minus signs are 
acquired from the delta function in
\begin{equation}
\nonumber
\langle\tilde\delta^{(1)}(\vec k_i)\tilde\delta^{(1)}(\vec k_j)\rangle=P(k_i)\delta^3_D(\vec k_i+\vec k_j) .
\end{equation}
Figure \ref{kurtfig} recreates Figure \ref{skewfig} for the smoothed kurtosis. 
Again we find that it is nearly impossible to distinguish a large change in
$\varpi_0$ from a small change in $\Omega_m$.

\end{appendix}

\end{document}